\documentclass[useAMS,usenatbib]{mn2e}
\input{epsf}
\usepackage{amssymb}
\usepackage{amsmath}
\usepackage{natbib}
\usepackage{color}
\usepackage{journals}
\usepackage{graphicx}
\usepackage{caption}
\usepackage{subcaption}

\newbox\grsign \setbox\grsign=\hbox{$>$} \newdimen\grdimen \grdimen=\ht\grsign
\newbox\simlessbox \newbox\simgreatbox
\setbox\simgreatbox=\hbox{\raise.5ex\hbox{$>$}\llap
     {\lower.5ex\hbox{$\sim$}}}\ht1=\grdimen\dp1=0pt
\setbox\simlessbox=\hbox{\raise.5ex\hbox{$<$}\llap 
     {\lower.5ex\hbox{$\sim$}}}\ht2=\grdimen\dp2=0pt

\newcommand{\hMpc}{{\ifmmode{h^{-1}{\rm Mpc}}\else{$h^{-1}$Mpc }\fi}}
\newcommand{\hGpc}{{\ifmmode{h^{-1}{\rm Gpc}}\else{$h^{-1}$Gpc }\fi}}
\newcommand{\hkpc}{{\ifmmode{h^{-1}{\rm kpc}}\else{$h^{-1}$kpc }\fi}}
\newcommand{\hMsun}{{\ifmmode{h^{-1}{\rm {M_{\odot}}}}\else{$h^{-1}{\rm{M_{\odot}}}$}\fi}}
\newcommand{\Msun}{{\ifmmode{{\rm {M_{\odot}}}}\else{${\rm{M_{\odot}}}$}\fi}}

\title[Two-population model of type Ia supernovae]{Two-population Bayesian hierarchical model of type Ia supernovae}
\author[R. Wojtak]{Rados{\l}aw~Wojtak$^{1}$\thanks{E-mail: radek.wojtak@nbi.ku.dk}, Jens Hjorth$^{1}$, Jacob Osman Hjortlund$^{1}$ \\
$^{1}$DARK, Niels Bohr Institute, University of Copenhagen, Jagtvej 128, 2200 Copenhagen, Denmark  \\
}
\begin{document}

\maketitle

\begin{abstract}
The currently used standardisation of type Ia supernovae results in Hubble residuals whose physical origin is unaccounted for. Here, we present a complete physical interpretation of the Hubble residuals based on a novel Bayesian hierarchical model of type Ia supernovae in which latent variables describing intrinsic and extrinsic (dust related) supernova properties originate from two supernova populations. Fitting the model to SALT2 light curve parameters of supernovae in the Hubble flow 
we find strong ($4\sigma$) evidence for the 
presence of two overlapping, but distinct, populations differentiated primarily by their mean SALT2 shape parameter (stretch) $x_{1}$. Supernovae from the population with predominantly slow decliners (higher average $x_{1}$) are found to be intrinsically bluer (mean SALT2 colour $c\approx-0.11$) 
and twice as reddened by dust (mean reddening $E(B-V)\approx 0.10$) than those from the other population which is dominated by fast decliners (lower average $x_{1}$) with $c\approx-0.04$ and $E(B-V)\approx 0.05$. The inferred extinction coefficient $R_{\rm B}$ in both supernova populations 
follows a broad distribution (scatter $0.9$) with a mean of $4.1$, 
which coincides closely with the value associated with mean extinction law in the Milky Way. We also find that the supernova data favour a peaked (two-tailed) distribution 
of selective extinction $E(B-V)$ over the commonly adopted exponential model. Our approach provides a complete explanation of the distribution
of supernova light curve parameters in terms of extinction properties and the above-mentioned differences 
between the two populations, without the need for introducing any intrinsic scatter.

\end{abstract}

\begin{keywords}
cosmology: observations -- cosmology: distance scale -- transients: supernovae -- methods: statistical
\end{keywords}

\section{Introduction}

Type Ia supernovae serve as one of the primary probes of cosmological models. 
They were a cornerstone of the discovery of cosmic acceleration \citep{Rie1998,Per1999} 
followed by establishing the standard $\Lambda$CDM cosmological model. Today, they are used to produce some of the leading data sets to constrain the dark energy equation of state \citep{Brout2022,Jones2019} and play a key role in 
high precision determinations of the Hubble constant \citep{Riess2022,Freed2019}. Type Ia supernovae 
will also be one of the major drivers of future cosmological tests based on observations from massive 
sky surveys conducted with next-generation telescopes such as the Vera C. Rubin Observatory \citep{LSST2009} and the {\em Nancy Grace Roman Space Telescope} \citep{Wang2022}.

The key element of type Ia supernova analysis is a process of standardisation whose goal is 
to improve the precision of distance measurements by means of applying empirical corrections 
based on observationally determined correlations between supernova apparent magnitudes 
and selected properties of their light curves. The standard approach accounts for corrections 
in two well measurable observables: the rest-frame width of the light curve (hereafter stretch parameter) 
and the apparent colour at a fixed phase of the light curve (typically at the peak of the light curve). 
The corrections are purely empirical with all relevant coefficients measured directly from data. 
This strategy was proposed a few decades ago by \citet{Tripp1998} and since then it has been 
used in virtually all cosmological analyses with type Ia supernovae. 

The stretch 
correction was first established by \citet{Phillips1999}. \citet{Kasen2007} showed that the 
correction most likely reflects an underlying relation between the the mass of radioactive $^{56}\rm{Ni}$ synthesised in the supernova explosion, which is the primary determinant of the supernova peak bolometric luminosity \citep{Arnett1982}, and broad absorption features developed after the peak and dominating in the $B$-band, although a possible modulation by the total ejected mass and its relation to the progenitor may complicate this picture \citep{Scalzo2014,Scalzo2014a}. 

The 
physical nature of the colour correction is quite different. The correction is a hybrid term which mixes possible effects related to supernova intrinsic colours, which may vary across supernovae and their progenitors, and dust reddening which is naturally expected to vary across supernova host galaxies and local environments. Many attempts to disentangle supernova intrinsic colours from the effect of dust have led to independent estimates of the extinction coefficient $R_{\rm B}=A_{\rm B}/E(B-V)$, i.e. the ratio of total $A_{\rm B}$ to selective $E(B-V)$ dust extinction in the $B$-band. The results point to extremely low values \citep[$R_{\rm B}\lesssim 3$; see e.g.][]{Nobili2008,Cikota2016,Wang2008,Hicken2009} relative to average $R_{\rm B}=4.3$ measured in the Milky Way \citep[][]{Schlafly2016} and $R_{\rm B}=3.8$ estimated in external galaxies \citep[][]{Finkelman2010}. This apparent discrepancy is a long-lasting problem and it signifies that either there are special low-extinction conditions operating solely in supernova sight lines \citep[see e.g.][]{Goobar2008,Bulla2018} or the measurements of extinction from supernova observations need to be revised with more accurate models \citep{Mandel2017}.

The standard two-parameter correction of supernova peak magnitudes \citep[hereafter the Tripp calibration after][]{Tripp1998} gives rise to irreducible intrinsic scatter in the Hubble residuals. The intrinsic scatter is a well measured property of the supernova Hubble diagrams and is typically found to be $\gtrsim0.1$~mag with minor improvements from applying phenomenological second-order corrections such as a step function in the host stellar mass \citep{Kelly2010,Scolnic2018,Jones2019,Smith2020} or the local specific star formation rate \citep{Rigault2020}. Inevitably, it is also a standard nuisance parameter in cosmological analyses of type Ia supernovae where it is one of the main sources of the total uncertainty in supernova distance moduli.

The persistence of a non-vanishing intrinsic scatter could be regarded as a sign that the Tripp calibration 
provides an incomplete framework for modelling type Ia supernovae. The argument is two-fold. Firstly, 
intrinsic scatter in its simplest form does not account for a range of second-order effects in the distribution 
of the Hubble residuals. Examples include an excess of positive residuals in red supernovae from low-mass host galaxies and a 
trend of the intrinsic scatter increasing with supernova colour \citep[larger scatter in redder supernovae;][]{Brout2021,Popovic2021}. 
Substantial differences between the Hubble residual distributions are also apparent when comparing the calibration sample 
(host galaxies with distances calibrated with Cepheids) and the Hubble flow sample of type Ia supernovae used in the local 
determination of the Hubble constant \citep{Wojtak2022}. Secondly, intrinsic scatter as such is naturally expected 
to result from neglecting a range of latent variables. Understanding the physical origin of the Hubble residuals and 
the intrinsic scatter should be imperative not only for the reason of developing models which are more rooted in first 
principles, but also for the sake of eliminating potential biases in cosmological measurements arising from unaccounted for supernovae properties.

Cosmological biases may occur, for example, when one neglects the observationally permissible scenario in which type Ia 
supernovae originate from two populations of progenitors, tracing old and young stellar populations, respectively \citep{Rigault2020}. 
The bias in this case would be a direct consequence of ignoring the redshift dependence of the population weights regulated 
by the star formation history.

Recent developments of Bayesian hierarchical models of type Ia supernovae operating at the level of light curve parameters \citep[see e.g.][]{Mandel2017,Brout2021,Popovic2021} or spectral energy energy distributions as light curve fitters \citep{Mandel2022} have helped pin down the physical cause of a substantial fraction of the intrinsic scatter in supernova Hubble diagrams. 
With physically motivated priors for dust reddening, models can disentangle in a probabilistic way the effect of dust from supernova intrinsic colours. 
Analyses of several different supernova samples concluded consistently that excessive and asymmetric Hubble residuals 
in red supernovae can be attributed to a wide range of extinction coefficients \citep{Thorp2021,Brout2021}. The 
implied distributions of $R_{\rm B}$ overlap substantially with those known from the Milky Way 
\citep{Fitzpatrick1999,Schlafly2016}, although they tend to peak at slightly lower values: $R_{\rm B}\approx 3.8$ \citep{Mandel2017,Thorp2021} relative to $R_{\rm B}\approx4.3$ measured in the Milky Way \citep{Schlafly2016}. The implied dust model leads to a reduction of the intrinsic scatter by about 40 
per cent \citep{Thorp2021,Mandel2022}. \citet{Brout2021} showed that further reduction is possible when extinction 
properties are inferred independently in two bins of supernova host galaxies split by the stellar mass with respect 
the transition mass of the mass step correction, i.e. $10^{10}M_{\odot}$ \citep[although see][]{Uddin2020}. This analysis implies very low values of the 
extinction coefficient in more massive host galaxies with respect to the Milky Way 
\citep[$R_{\rm B}\approx 2.5$ compared to $R_{\rm B}\approx 4.3$ from][]{Schlafly2016}. This result begs the question what physical mechanisms can 
form dust with this extinction property and why they are efficient only in host galaxies with stellar masses larger than $10^{10}M_{\odot}$.

One limitation of the currently proposed Bayesian hierarchical models of type Ia supernovae is the assumption
that type Ia supernovae form a single population whose intrinsic and extrinsic properties are drawn from unimodal 
prior distributions. This assumption can hardly be reconciled with observations which provide evidence for the existence 
of two supernova populations distinguished by the decline rate of their light curves and the stellar age of their environments 
\citep{Rigault2013,Rigault2020,Maoz2014}. Fast declining supernovae are typically found in old stellar populations, while 
slowly declining supernovae originate in young star forming environments \citep{Sullivan2006}. Both supernova populations exhibit slightly different normalisation 
of their Hubble diagrams and when taken into account in the standardisation, this property can effectively reduce the 
intrinsic scatter \citep{Rigault2020}. This suggests that the notion of two supernova populations should be regarded 
as a potentially important element of a model explaining the Hubble residuals and perhaps a missing element of the 
present Bayesian hierarchical models. The two observationally distinguished populations linked to young and old 
stellar environments likely originate from different progenitors channels.
Based on an argument of time scales between the formation of a binary system and the supernova explosion, it is tempting 
to associate progenitors of supernovae in young (old) stellar populations with single-degenerate (double-degenerate) 
systems \citep{Maoz2014}. The two progenitor channels would manifest themselves as prompt (short time scale) and delayed 
(long time scale) supernovae whose rates follow the star formation history (prompt) or are lagging behind (delayed) 
with the delay time following the distribution $\propto t^{-1}$ expected for double degenerate progenitors. Measurements 
of the type Ia supernova rate as a function of redshift or galaxy properties are consistent with the presence of both prompt and delayed supernova 
populations \citep{Scannapieco2005,Mannucci2006,Rodney2014,Andersen2018}. Although the two-channel scenario is commonly 
accepted as a framework for studying supernova progenitors, we emphasise that its status should regarded as a working hypothesis 
rather than a fully confirmed theory \citep{Livio2018}.

The goal of our study is to incorporate the notion of two supernova populations (with no prior assumptions on their physical properties) in a Bayesian hierarchical model 
of type Ia supernova light curve parameters. We aim to demonstrate that including this observationally motivated assumption 
enables us to fully account for the intrinsic scatter resulting from the Tripp calibration in terms of effects of dust and differences 
between the physical properties of two supernova populations separated probabilistically by the new model. Using a data-driven 
approach we also improve the prior for the distribution of dust reddening adopted in previous studies. This prior distribution is the key element of disentangling dust reddening from supernova intrinsic colour in the Bayesian hierarchical modelling.

An important motivation for developing the model is to prepare the ground for a physical interpretation of the recently found intrinsic tension in the supernova sector of the local determination of the Hubble constant with distances calibrated with Cepheids \citep{Wojtak2022}. The tension arises as a discrepancy between the colour corrections of type Ia supernovae in the calibration sample (host galaxies with independently observed Cepheids) and the Hubble flow, with (only) the former being entirely consistent with a typical extinction correction in the Milky Way. This represents an intrinsic anomaly of the Tripp calibration, which assumes universality of the colour correction across all supernova samples, and thus a potential source of unaccounted systematic errors in the local measurement of the Hubble constant from the Supernovae and $H_{0}$ for the Dark Energy Equation of State (SH0ES) program \citep{Riess2016,Riess2019,Riess2021}. Since host galaxies in the calibration sample are selected as late type galaxies containing observable Cepheids, it is natural to investigate if the different supernova colour corrections in these samples reflect a different mixture of supernova populations (perhaps dominated by one of them) and extinction in the underlying young stellar environments in the calibration sample vs.\ the host galaxies of the Hubble flow. Although selecting similar late-type galaxies in the Hubble flow as in the calibration sample has a negligible impact in the Hubble constant determination \citep{Riess2022}, it is imperative to check whether other approaches can corroborate this conclusion or not. Unlike the strategy based on matching galaxy properties on the two rungs of the distance ladder, Bayesian modelling of supernova light curve parameters is capable of measuring probabilistic properties of extinction in sight lines towards observed supernovae. We expect that the model developed in our work will shed more light on this problem and its impact on the Hubble constant determination.

The outline of the paper is as follows. In section 2 we describe the model including an implementation of the two-population 
assumption, the adopted hyperpriors and the related hyperparameters as well as the inference method. The supernova 
data and the results of fitting several different versions of the model are presented in section 3. In this section we also 
demonstrate the completeness and accuracy of the final model in terms of accounting for supernova Hubble residuals 
from the Tripp calibration. In section 4 we discuss the implied intrinsic properties of supernovae and extrinsic properties 
of dust in the context of ongoing type Ia supernova studies. We summarise our findings in section 5.

\section{Model}

\subsection{Rationale}

The multi-band light curve of a normal type Ia supernova can be effectively described by three parameters: 
the peak magnitude, the rest-frame width of the light curve, and the apparent colour. These parameters can be measured directly from observations by fitting empirical models of multi-band light curves generated from spectral templates obtained for a large number of
supernovae. One can think of light curve parameters as a minimum set of observables describing a normal type Ia supernova 
light curve without loss of information (maximum data compression). In this study, we use parameters obtained with the SALT2 light 
curve fitter \citep{Bet2014}: light curve amplitude quantified by the apparent $B$-band peak magnitude in the supernova rest frame 
$m_{\rm B}$, the dimensionless parameter $x_{1}$ describing the light curve stretch, and the colour parameter $c$ describing the restframe $B-V$ colour at restframe $B$-band peak. 

Light curve parameters only provide a phenomenological description of the observed supernova light curves. Their relation to the corresponding intrinsic (e.g. intrinsic supernova colour) and extrinsic (e.g. dust reddening and extinction) physical parameters is highly degenerate due to the fact that the number of relevant latent variables is typically larger than the number of light curve parameters. 
This problem is immediately apparent taking as an example the supernova colour: while a single parameter suffices to describe the apparent supernova colours, the actual physical 
parameters needed to explain them include intrinsic colours related to the physics of supernova explosions and initial conditions, 
and reddening due to dust in the host galaxy. Although most of the physical parameters cannot be directly inferred from 
observed light curves due to the above-mentioned degeneracy, they do shape the distribution of directly measured light curve parameters. Therefore, modelling the distribution 
of light curve parameters can be used to determine statistical properties of latent physical parameters for a given supernova population. This inference is a standard problem which can be handled using Bayesian hierarchical modelling.

As a starting point for Bayesian hierarchical modelling we formulate relations between observables and latent variables. 
We adopt a commonly used model which enables us to separate intrinsic supernova properties from extrinsic effects related to dust reddening 
and extinction \citep[see e.g.][]{Mandel2017,Brout2021}. The model quantifies the intrinsic properties in terms of the absolute peak 
magnitude $M_{\rm B}$, the stretch parameter $X_{1}$ (assumed to be equal to the stretch parameter $x_{1}$ from SALT2 and distinguished from it only for the sake of concise mathematical formulation below), 
the intrinsic $B-V$ colour $c_{\rm int}$ at the light curve maximum as well as empirical coefficients of possible relations between $m_{\rm B}$ and $\{X_{1}$, $c_{\rm int}\}$. The intrinsic parameters only provide a phenomenological description of physical processes behind the observed supernova light curves. However, they can be directly linked to initial conditions 
of type Ia supernovae given a complete physical model of supernova explosions. The extrinsic latent variables include dust reddening 
$E(B-V)$ and extinction coefficient $R_{\rm B}$. 

The model relating the latent variables to the measured light curve parameters is given by the following equations:
\begin{equation}
\begin{aligned}
m_{\rm B} & =M_{\rm B}-\alpha X_{1} +\beta c_{\rm int}+\mu(z)+R_{\rm B}E(B-V) \\
x_{1} &=X_{1} \\
c&=c_{\rm int}+E(B-V),
\end{aligned}
\label{observables}
\end{equation}
where $\mu(z)$ is the distance  modulus and $z$ is the CMB rest frame redshift of the supernova. It can be 
thought of as the simplest generalisation of the Tripp calibration \citep{Tripp1998} which can be recovered when the effect of extinction and intrinsic colour correction are indistinguishable, i.e. $\beta=R_{\rm B}$. For the sake of simplicity, we will hereafter refer to all variables and coefficients on the right hand side of the equations as a latent variable vector $\pmb{\phi}$, 
i.e. $\pmb{\phi}=\{M_{\rm B},X_{1},c_{\rm int},E(B-V),\alpha,\beta,R_{\rm B}\}$, and to those on the left hand side as observables 
$\pmb{\xi}=\{m_{\rm B},x_{1},c\}$.

\subsection{Two-population model}

Observations provide evidence that type Ia supernovae with fast or slowly declining light curves (low or high stretch parameter) originate from passive and star-forming environments, respectively \citep{Sullivan2006,Rigault2013,Rigault2020,Larison2023}. The 
corresponding supernova populations can be identified probabilistically as two components in the distribution of stretch parameters. A bimodal distribution of stretch parameter signifying the presence of two supernova populations is apparent in many low-redshift 
supernova compilations \citep{Scolnic2016,Scolnic2018,Dhawan2022} and volume limited samples at high redshifts \citep{Nicolas2021}. We 
incorporate this observational fact in our model by introducing two separate supernova populations for which prior 
distributions of latent variables can differ. We assume that possible differences between prior distributions of the two populations may occur 
in variables associated with intrinsic and extrinsic properties. Differences between intrinsic properties may reflect two sets of initial conditions in the two supernova populations, which may in turn be related to the progenitor channels. Analogous differences in the dust 
sector can arise from diverse conditions of supernova local environments and lines of sight. They can potentially reflect diverse dust properties in star-forming and 
passive environments traced by the two supernova populations.

We employ a two-population model in a probabilistic way by introducing a prior probability $w$ that a given supernova originates 
from one of the two supernova populations. The prior probability regulates the ratios of supernova populations in a given sample 
so that the probability of observing a supernova with latent (physical) parameters $\pmb{\phi}$ is given by
\begin{equation}
p(\pmb{\phi})=wp_{\rm prior\,1}(\pmb{\phi})+(1-w)p_{\rm prior\,2}(\pmb{\phi}),
\label{prior}
\end{equation}
where $p_{\rm prior\,i}$ is the prior probability distribution of latent variables in the $i$-th supernova population. Once the prior probabilities 
in the two populations are known one can determine the distribution of observables $\pmb{\xi}_{0}$ as
\begin{equation}
p(\pmb{\xi}_{0})=\int \delta(\pmb{\xi}_{0}-\pmb{\xi}(\pmb{\phi})) [wp_{\rm prior\,1}(\pmb{\phi})+(1-w)p_{\rm prior\,2}(\pmb{\phi})]\textrm{d}\pmb{\phi},
\label{margin_prior}
\end{equation}
where $\pmb{\xi}(\pmb{\phi})$ is given by eqs.~(\ref{observables}) and the integral over latent variables can be effectively computed 
in a Monte Carlo way by sampling from the prior distributions of the two supernova populations. We will use this equation to show a 
quantitative interpretation of well studied residuals in Hubble diagrams of type Ia supernova standardised with the Tripp calibration \citep{Tripp1998}.

We constrain the prior weight $w$ and the properties of the resulting populations solely from the supernova data. 
The two populations are disentangled probabilistically 
through correlated signatures of two distinct components in the 
distributions of the latent variables. As shown by \citet{Rigault2020} and \citet{Nicolas2021}, type Ia supernovae with slowly or fast declining light curves 
(positive or negative $x_{1}$) can be differentiated not only by the local specific star formation rate of their environments but also 
as two separable components of the stretch parameter distribution. It is perhaps not surprising that the two supernova populations 
emerging in our model are primarily distinguished by their mean stretch parameters and represent fast and slowly declining 
supernovae. What is unique in our model is that it enables us to quantify which remaining intrinsic and extrinsic properties are 
common to the both populations and which are different.

\subsection{Priors}

We assume that all latent variables except for $\alpha$, $\beta$ and $E(B-V)$ follow independent Gaussian distributions. The corresponding 
means and standard deviations are free parameters (hyperparameters) to be constrained by observations. We assume that 
both $\alpha$ and $\beta$ are single-valued parameters with the corresponding distributions given by $\delta$ functions. 

\begin{figure}
	\centering
	\includegraphics[width=\linewidth]{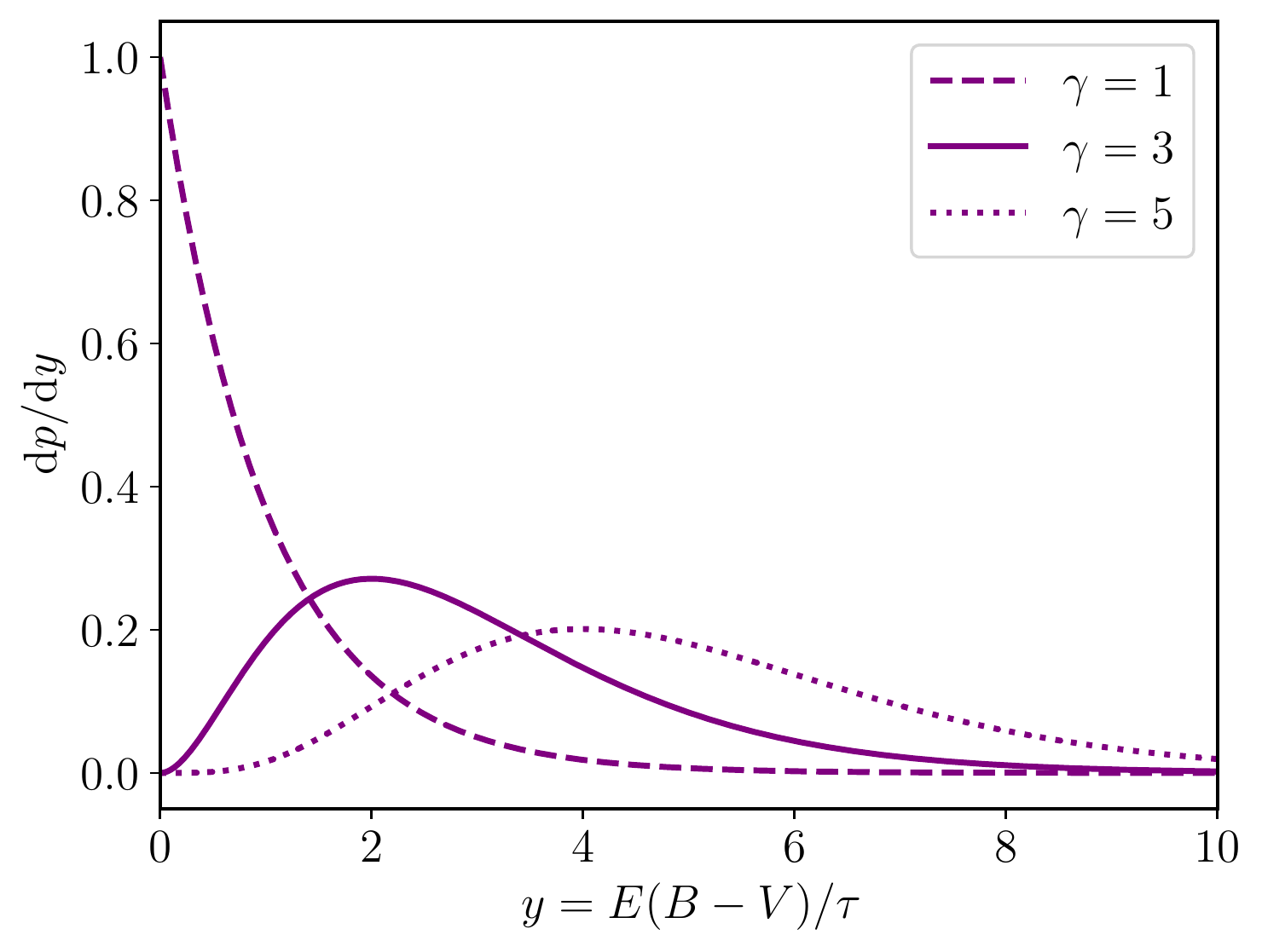}
	\caption{The prior probability distribution of $y=E(B-V)/\tau$ for a range of shape parameter values, plotted up to the adopted upper limit $y_{\rm max}=10$.
	}
	\label{EBV_prior}
\end{figure}

The selective extinction $E(B-V)$ is a positively defined variable. This is a strict physical condition which narrows down a range of possible prior distributions. 
In our model we adopt a flexible two-parameter family of probability distributions given by the gamma distribution:
\begin{equation}
p_{\rm prior}(y=E(B-V)/\tau)=\frac{y^{\gamma-1}\exp(-y)}{\Gamma(\gamma)}.
\end{equation}
The assumed class of probability distributions is the maximum information entropy solution (the most likely function) for a positively defined variable 
subject to constrained mean values of the variable and its logarithmic counterpart. The free parameters describe the shape of the distribution ($\alpha$) 
and a characteristic scale of the most probable values ($\tau$). The gamma distribution reduces to an exponential model when $\gamma=1$. All 
distributions with $\gamma>1$ are peaked at $y>0$ (two-tailed distributions), while the exponential case is represented by one-tailed distributions 
with maximum at $y=0$. In order to control the accuracy of numerical integration 
over $y=E(B-V)/\tau$ in the likelihood evaluation (see the following sections and Appendix A), we cut off the distribution's upper tail at $y_{\rm max}=10$ and use the resulting renormalized truncated distributions. Figure~\ref{EBV_prior} shows examples of the prior distribution plotted between $y=0$ and $y_{\rm max}$ for a range of shape parameter values.

Our choice of the prior distribution of dust reddening generalises the commonly used exponential model which is adopted in virtually all 
Bayesian hierarchical models from the literature \citep[see e.g.][]{Mandel2017,Mandel2022,Popovic2021,Brout2021}. The generalisation 
can be justified in a two-fold way. From a statistical point of view, the choice 
of maximum entropy solution and the corresponding parameterization should be adjusted to the constraining power of the data in a way that there is 
adequate correspondence between the information capacity of the model and the actual information in the data. Unlike previous studies, we adopt equal numbers 
of hyperparameters (degrees of freedom) in each of the two prior distributions relevant in modelling the distribution of apparent colours: $p(c_{\rm int})$ and $p(\rm{E(B-V)})$. As we shall see in the following section, the shape parameter $\gamma$ of the prior 
distribution of dust reddening is not a redundant parameter and it appears to be equally well constrained by the supernova data as other hyperparameters. 
The second argument involves basic considerations of possible geometric configurations of supernovae and intervening dust in a typical supernova host galaxy. 
For a randomly distributed dust clouds and supernova locations, it is natural to expect that the most probable configuration occurs for some finite column 
density of dust and the corresponding reddening. This configuration can be easily reconciled with the gamma distribution model for which the maximum 
probability position is directly related to the shape parameter $\gamma$ (the larger $\gamma$, the larger shift of the distribution's peak). The exponential 
model does not have this flexibility and sets the maximum probability at $E(B-V)=0$.

For the remaining parameter of the dust sector, i.e., the extinction coefficient $R_{\rm B}$, we assume a Gaussian prior distribution. We do not apply any 
truncation which could exclude $R_{\rm B}$ values which are not allowed on the the grounds of theoretical dust models. However, as we shall see in the following 
section, only low $2\sigma$ limits of our best fit models are comparable to the lower limit expected for Rayleigh scattering \citep[$R_{\rm B}=2.2$;][]{Draine2003} 
and well above $R_{\rm B}=1.5$ adopted as a more liberal limit in analyses similar to ours \citep{Brout2021,Thorp2021}. This leaves only 1--2 per cent 
of the lowest $R_{\rm B}$ values below theoretical limits. This fraction is sufficiently small to conclude that neglecting any truncation in the prior distribution 
of $R_{\rm B}$ has a negligible impact on the final results.

The distance modulus latent variable in eqs.~(\ref{observables}) is assumed to have a Gaussian distribution. The mean value is computed for the Planck 
cosmological model \citep{Planck2020_cosmo} with the Hubble constant renormalised to $H_{0}=70\,{\rm km}\,{\rm s}^{-1}{\rm Mpc}^{-1}$. Since we fit 
our model to supernovae in a redshift range of the Hubble flow (see more details in the following section), the assumption of a cosmological model does 
not have any noticeable impact on the final results and the adopted Hubble constant fixes a reference value for the absolute magnitudes. The main source 
of uncertainties in $\mu(z)$ comes from unconstrained peculiar velocities in the local volume. This results in the following standard deviation
\begin{equation}
\sigma_{\mu}=\frac{5}{\ln 10}\frac{\sigma_{v}}{cz},
\end{equation}
where $\sigma_{v}=250\,{\rm km}\,{\rm s}^{-1}$ is a commonly adopted value of the line-of-sight peculiar velocity dispersion assumed in cosmological fits with type Ia 
supernovae \citep{Scolnic2018} and equal to an upper limit of scatter in peculiar velocities with respect to the linear velocity field \citep{Carrick2015}.

All hyperparameter names are listed in Table~\ref{hyperparameters}. We use an intuitive notation in which $\widehat{X}$ and $\sigma_{X}$ are used to refer to 
the mean and standard deviation of the corresponding latent variable $X$. For the sake of simplicity, we keep the same names of single-value 
parameter as their corresponding latent variables. We use a distinct name for the standard deviation of $M_{\rm B}$. This parameter 
is a direct Bayesian counterpart of the commonly used `intrinsic scatter' which quantifies excessive (unaccounted for by the model) 
residuals in supernova Hubble diagrams. For this reason, we label it $\sigma_{\rm int}$.

\begin{table}
\begin{center}
\begin{tabular}{lll}
latent & prior & hyperparameters \\
variable &  & [hyperpriors] \\
\hline
$M_{\rm B}$ & $\mathcal{G}(M_{\rm B};\widehat{M_{\rm B}},\sigma_{\rm int})$ & $\widehat{M_{\rm B}}[-\infty,+\infty]$, $\sigma_{\rm int}[0,+\infty]$ \\
$X_{1}$ & $\mathcal{G}(X_{1};\widehat{x_{1}},\sigma_{x_{1}})$ & $\widehat{x_{1}}[-\infty,+\infty]$, $\sigma_{x_{1}}[0,+\infty]$ \\
$c_{\rm int}$ & $\mathcal{G}(c_{\rm int};\widehat{c_{\rm int}},\sigma_{c_{\rm int}})$ & $\widehat{c_{\rm int}}[-\infty,+\infty]$, $\sigma_{c_{\rm int}}[0,+\infty]$ \\
$\alpha$ & $\delta(x-\alpha)$ & $\alpha[-\infty,+\infty]$ \\
$\beta$ & $\delta(x-\beta)$ & $\beta[1,5]$ \\
$E(B-V)$ & $\propto x^{\gamma-1}\exp(-x/\tau)$ & $\tau[0,+\infty]$,$\gamma[1,7]$ \\
$R_{\rm B}$ & $\mathcal{G}(R_{\rm B};\widehat{R_{\rm B}},\sigma_{R_{\rm B}})$ & $\widehat{R_{\rm B}}[1.5,6]$,$\sigma_{R_{\rm B}}[0,1.5]$ \\
\end{tabular}
\caption{Prior probability distributions of all latent variables and the corresponding hyperparameters. Each 
set of hyperparameters can be different in two supernova populations of the model. The last column 
lists all hyperparameters associated with a given prior distribution and the corresponding ranges 
of allowed values (flat hyperprios). $\mathcal{G}(x;\mu,\sigma)$ denotes a normal distribution with 
mean $\mu$ and standard deviation $\sigma$, and $\delta(x-x_{0})$ is the Dirac delta distribution.
}\label{hyperparameters}
\end{center}
\end{table}

\subsection{Likelihood and inference}

We constrain free hyperparameters by fitting the model to supernova data in the form of measured light curve parameters $\pmb{\xi_{\rm obs\,i}}$, 
the corresponding covariance matrix $\mathsf{C_{\rm obs,i}}$ and redshift $z_{i}$. The likelihood used in our analysis is given by the following 
equation
\begin{equation}
L \propto \prod_{i}^{N}p(\{\pmb{\Theta},w\}|\pmb{\xi_{\rm obs\,i}},\mathsf{C_{\rm obs\,i}},z_{i}),
\label{likelihood1}
\end{equation}
where $\pmb{\Theta}$ is a vector of all hyperparameters in both supernova populations, $N$ is the number of supernovae and probability distribution 
$p$ is calculated by marginalising the product of the Gaussian probability distribution accounting for measurement uncertainties and the prior 
probability given by eq.~(\ref{prior}) over all latent variables, i.e.
\begin{equation}
p(\{\pmb{\Theta},w\}|\pmb{\xi_{\rm obs\,i}},\mathsf{C_{\rm obs\,i}},z_{i})=\int \mathcal{G}[\pmb{\xi}(\pmb{\phi});\pmb{\xi_{\rm obs\,i}},\mathsf{C_{\rm obs\,i}}]
p_{\rm prior}(\pmb{\phi})\textrm{d}\pmb{\phi}.
\label{likelihood2}
\end{equation}
Integration over the latent variables which have Gaussian prior distributions and occur as linear terms in the model of observables, i.e. 
$\{M_{\rm B}, x_{1},c_{\rm int}, \mu, R_{\rm B}\}$, results in a sum of two Gaussian distributions weighted by $w$ and $(1-w)$. 
Integration over the remaining variable $E(B-V)$ does not have an analytical solution and thus it is carried out numerically. Therefore, 
each evaluation of the likelihood involves $N$ numerical integrations over $E(B-V)$. We outline all necessary 
details including explicit forms of the covariance matrices resulting from the analytical part of the marginalisation in Appendix A. 

We compute best fit parameters by means of integrating the posterior probability using a \textit{Monte Carlo Markov Chain} technique 
implemented in the \textit{emcee} code \citep{emcee}. For most of the hyperparameters we employ rather unrestrictive priors. In order 
to reduce the level of a lower tail of the prior distribution at $R_{\rm B}\lesssim 1.5$ to less than $2\sigma$, we restrict the range of 
$\sigma_{R_{\rm B}}$ to $[0,1.5]$. We also use finite limits of hyperpriors for $\widehat{R_{\rm B}}$, $\beta$ and $\gamma$. For the mean extinction coefficient $\widehat{R_{\rm B}}$, we adopt the bounds given by a conservative theoretical lower limit of $1.5$ \citep{Brout2021,Thorp2021} and a maximum value of extinction ($R_{\rm B}\approx 6$) measured in the Milky Way by \citet{Fitzpatrick2007}. Exact limits of all hyperpriors are provided in Table~\ref{hyperparameters}. In all cases, the adopted hyperpriors are wider than the lower and upper limits inferred from the likelihood and the main reason of using them is to stabilise convergence of the chains. For the same reason, we restrict the range 
of parameter $w$ to $[0.1,0.9]$. This prevents the chains from populating models with $w\approx 0$ or $w\approx 1$ where hyperparameters 
of a zero-weighted supernova population looses any constraints and its hyperparameters can diverge from the equilibrium solution. As we shall 
see in the following section, the adopted prior for $w$ is wider than the actual limits from the likelihood and thus it has no impact 
on the final results. Unless explicitly stated, best-fit parameters are provided in the form of posterior means and errors are computed as 
16th and 84th percentiles of the marginalised probability distributions. The $1\sigma$ and $2\sigma$ confidence contours in all figures contain 
68 and 95 per cent of the corresponding 2-dimensional marginalised probability distributions.

\section{Observational constraints}

\subsection{Observations}

We use type Ia supernova light curve parameters from the \textit{SuperCal} compilation \citep{Scolnic2015}. The supernova sample was compiled from several different surveys using a consistent photometric calibration based on Pan-STARRS observations. Supernovae in the Hubble flow 
were used in measurements of the Hubble constant presnted in \citet{Riess2016,Riess2019,Riess2021} 
based on consecutive improvements of distance anchors. The compilation provides best fit light curve parameters of the SALT2 model \citep{Bet2014}: 
the apparent $B$-band peak magnitude in the supernova rest frame $m_{\rm B}$, dimensionless parameter $x_{1}$ describing the light curve shape 
(stretch parameter) and colour parameter $c$ describing the observed $B-V$ colour in the supernova rest frame, as well as the corresponding covariance 
matrices which are essential for accurate evaluation of the likelihood.

In order to avoid possible biases related to selection effects of the surveys included in the supernova sample, we focus on relatively low-redshift supernovae for which these effects are minimised. Observational biases become non-negligible 
at high redshifts and the commonly used strategy to remove them is to simulate observations given a model of light curves and redshift-evolution of light curve parameters \citep[see e.g.][]{Scolnic2018}. The current simulations involve the standard 
supernova calibration based on the Tripp calibration and as such they cannot be implemented 
in out modelling in a self-consistent way. Another complication 
arises from the fact that the prior weight $w$ can be a function of redshift. Including this effect in the analysis of high-redshift 
data would require a full forward modelling of all components (cosmological model, Bayesian model of type Ia supernovae and survey 
selections) in order to control possible degeneracies. Conversely, considering low-redshift supernovae allows us to assume a constant weight $w$ and to obtain constraints on supernova properties independently of cosmology and selection effects.

We select supernovae in the Hubble flow at redshifts between $0.023$ and $0.15$. We also apply additional selection criteria from \citet{Riess2016} regarding the allowed allowed range of light curve parameters and the quality of light curve fits. Specifically, we include 
supernovae with the colour parameter $|c|<0.3$ and the stretch parameter $|x_{1}|<3$. These cuts eliminate a relatively small 
group of outliers for which the accuracy of interpolation built in the light curve SALT2 model is clearly worse than for normal type Ia supernovae. 
In addition, we omit supernovae with poor quality indicated by at least one of the following conditions: $fitprob>0.001$, error in 
the stretch parameter larger than $1.5$, error in the peak time larger than 2 days or error in the corrected magnitude 
(approximated by the Tripp formula) larger than $0.2$~mag. This leaves us with $222$ supernovae with high quality measurements 
of light curve parameters.

In order to test the impact of recent updates in the light curve fits, calibrations \citep{BroutTaylor2022} 
and redshift estimations \citep{Carr2022}, we repeat 
our analysis using supernova data from the \textit{Pantheon+} catalogue \citep{Brout2022}. We use 156 supernovae which 
overlap with our \textit{SuperCal} sample. For supernovae with multiple light curve parameter measurements from different surveys (duplicates), we combine all independent results and compute single-measurement equivalents. Combining duplicates is necessary in order to avoid artificial weights which could otherwise bias the underlying distributions of light curve parameters. In these cases, 
the best-fit light curve parameters and covariance matrices are given by products of Gaussian probability distributions with the means and covariance matrices from the corresponding duplicates. 
In our analysis we omit bias corrections and systematic errors provided in the \textit{Pantheon+} catalogue. The bias correction is primarily driven by the dust model of \citep{Brout2021} and thus it is not applicable to data modelling whose goal is to constrain extinction properties. Similarly, systematic errors are estimated for model-dependent derived distance moduli and therefore cannot be included in our analysis in a self-consistent way.

As an independent data set we also use the Foundation DR1 supernova sample \citep{Foley2018,Jones2019}. Compared to the \textit{SuperCal} sample, 
this is a far more homogenous data set with all observations having been obtained with same instrument (Pan-STARRS) 
and a high accuracy of photometric calibration. The Foundation survey is a follow-up survey observing selected targets from other 
transient surveys. The targets are chosen as spectroscopically confirmed type Ia supernovae satisfying a number of secondary conditions. 
Therefore, one can expect that the resulting supernova sample is not fully unbiased and possible biases are related primarily to 
spectroscopic classification strategy. We omit several supernovae at redshift $z<0.015$, for which the constraining power is 
significantly reduced due to large uncertainties related to peculiar velocities, so that the final sample contains 174 supernovae with 
a maximum redshift of $0.11$. The Foundation DR1 catalogue provides best fit light curve parameters and covariance matrices based 
on the same light curve fitter as the \textit{SuperCal} supernova sample (SALT2). The Foundation sample is used as a comparison data set in our study. 
Unless explicitly stated, all constraints and the final results are based on the \textit{SuperCal} sample described above.

\subsection{Baseline model}

Fits to the supernova data unambiguously reveal the presence of two distinct supernova populations ($w=0$ and $w=1$ ruled with $4\sigma$ significance). It is also apparent that the main discriminator of the two populations is the stretch parameter $x_{1}$. The two populations identified probabilistically 
in the data are supernovae with two decline rates of their light curves: fast declining with $\widehat{x_{1}}\approx-1.3$ and slow declining 
with $\widehat{x_{1}}\approx 0.4$. We will hereafter refer to these two populations respectively as population 1 (pop1) and population 2 
(pop2) with the corresponding red (population 1) and blue (population 2) colour pallets in all figures.

\begin{table*}
\begin{center}
\begin{tabular}{lcccccccc}
\hline
model & \multicolumn{2}{c}{baseline} & \multicolumn{2}{c}{baseline+$\sigma_{\rm int}$} & \multicolumn{2}{c}{single population} &  \multicolumn{2}{c}{baseline} \\
\hline
supernova & \multicolumn{2}{c}{SuperCal} & \multicolumn{2}{c}{SuperCal} & \multicolumn{2}{c}{SuperCal} &  \multicolumn{2}{c}{Foundation}\\
sample & \multicolumn{2}{c}{$0.023<z<0.15$} & \multicolumn{2}{c}{$0.023<z<0.15$} & \multicolumn{2}{c}{$0.023<z<0.15$} &  \multicolumn{2}{c}{$0.015<z<0.11$}\\
\hline
 & SNe(pop1) & SNe(pop2) & SNe(pop1) & SNe(pop2) & \multicolumn{2}{c}{pop1=pop2} & SNe(pop1) & SNe(pop2)\\
 & fast declining & slowly declining & fast declining & slowly declining & \multicolumn{2}{c}{} & fast declining & slowly declining\\
 \hline
 & \\
 $\widehat{M_{\rm B}}$ & $ -19.49 ^{+ 0.05 }_{- 0.05 }$ & $ -19.40 ^{+ 0.07 }_{- 0.07 }$ & $-19.44 ^{+ 0.06 }_{- 0.07 }$ & $ -19.37 ^{+ 0.07 }_{- 0.08 }$ & \multicolumn{2}{c}{$ -19.383 ^{+ 0.096 }_{- 0.092 }$} & $ -19.44 ^{+ 0.09 }_{- 0.09 }$ & $ -19.33 ^{+ 0.08 }_{- 0.08 }$ \\
& \\
$\widehat{x_{1}}$ & $ -1.31 ^{+ 0.28 }_{- 0.24 }$ & $ 0.41 ^{+ 0.14 }_{- 0.13 }$ & $-1.20 ^{+ 0.43 }_{- 0.35 }$ & $ 0.46 ^{+ 0.18 }_{- 0.16 }$ & \multicolumn{2}{c}{$ -0.250 ^{+ 0.075 }_{- 0.075 }$} & $ -0.72 ^{+ 0.27 }_{- 0.27 }$ & $ 0.76 ^{+ 0.16 }_{- 0.13 }$ \\
& \\
$\sigma_{x_{1}}$ & $ 0.70 ^{+ 0.16 }_{- 0.15 }$ & $ 0.64 ^{+ 0.09 }_{- 0.09 }$ & $0.76 ^{+ 0.23 }_{- 0.19 }$ & $ 0.59 ^{+ 0.12 }_{- 0.13 }$ & \multicolumn{2}{c}{$ 1.093 ^{+ 0.055 }_{- 0.055 }$} & $ 0.88 ^{+ 0.15 }_{- 0.16 }$ & $ 0.52 ^{+ 0.10 }_{- 0.13 }$ \\
& \\
$\widehat{c_{\rm int}}$ & $ -0.035 ^{+ 0.034 }_{- 0.033 }$ & $ -0.112 ^{+ 0.033 }_{- 0.036 }$ & $-0.043 ^{+ 0.040 }_{- 0.042 }$ & $ -0.103 ^{+ 0.034 }_{- 0.038 }$ & \multicolumn{2}{c}{$ -0.103 ^{+ 0.039 }_{- 0.042 }$} & $ -0.114 ^{+ 0.037 }_{- 0.037 }$ & $ -0.133 ^{+ 0.031 }_{- 0.031 }$ \\
& \\
$\sigma_{c_{\rm int}}$ & $ 0.077 ^{+ 0.011 }_{- 0.011 }$ & $ 0.042 ^{+ 0.013 }_{- 0.013 }$ & $0.069 ^{+ 0.017 }_{- 0.019 }$ & $ 0.043 ^{+ 0.014 }_{- 0.015 }$ & \multicolumn{2}{c}{$ 0.047 ^{+ 0.016 }_{- 0.016 }$} & $ 0.056 ^{+ 0.022 }_{- 0.021 }$ & $ 0.034 ^{+ 0.021 }_{- 0.022 }$ \\
& \\
$\tau$ & $ 0.017 ^{+ 0.008 }_{- 0.009 }$ & $ 0.034 ^{+ 0.008 }_{- 0.008 }$ & $0.023 ^{+ 0.015 }_{- 0.015 }$ & $ 0.037 ^{+ 0.011 }_{- 0.010 }$ & \multicolumn{2}{c}{$ 0.035 ^{+ 0.008 }_{- 0.010 }$} & $ 0.041 ^{+ 0.012 }_{- 0.015 }$ & $ 0.034 ^{+ 0.012 }_{- 0.014 }$ \\
& \\
$\alpha$ & \multicolumn{2}{c}{$ 0.185 ^{+ 0.024 }_{- 0.025 }$} & \multicolumn{2}{c}{$ 0.171 ^{+ 0.025 }_{- 0.024 }$} & \multicolumn{2}{c}{$ 0.139 ^{+ 0.009 }_{- 0.008 }$} &  \multicolumn{2}{c}{$ 0.153 ^{+ 0.036 }_{- 0.036 }$} \\
& \\
$\beta$ & \multicolumn{2}{c}{$ 3.089 ^{+ 0.243 }_{- 0.238 }$} & \multicolumn{2}{c}{$ 3.174 ^{+ 0.275 }_{- 0.334 }$} & \multicolumn{2}{c}{$ 3.193 ^{+ 0.532 }_{- 0.478 }$} & \multicolumn{2}{c}{$ 3.116 ^{+ 0.251 }_{- 0.251 }$} \\
& \\
$\widehat{R_{\rm B}}$ & \multicolumn{2}{c}{$ 4.132 ^{+ 0.647 }_{- 0.591 }$} & \multicolumn{2}{c}{$ 3.964 ^{+ 0.725 }_{- 0.669 }$} & \multicolumn{2}{c}{$ 3.698 ^{+ 0.515 }_{- 0.497 }$} & \multicolumn{2}{c}{$ 3.789 ^{+ 0.551 }_{- 0.546 }$} \\
& \\
$\sigma_{R_{\rm B}}$ & \multicolumn{2}{c}{$ 0.946 ^{+ 0.313 }_{- 0.283 }$} & \multicolumn{2}{c}{$ 0.781 ^{+ 0.371 }_{- 0.350 }$} & \multicolumn{2}{c}{$ 0.751 ^{+ 0.343 }_{- 0.302 }$} & \multicolumn{2}{c}{$ 0.843 ^{+ 0.298 }_{- 0.265 }$} \\
& \\
$\gamma$ & \multicolumn{2}{c}{$ 3.231 ^{+ 1.236 }_{- 1.258 }$} & \multicolumn{2}{c}{$ 2.717 ^{+ 1.321 }_{- 1.276 }$} & \multicolumn{2}{c}{$ 3.129 ^{+ 1.293 }_{- 1.360 }$} & \multicolumn{2}{c}{$ 3.296 ^{+ 1.197 }_{- 1.234 }$} \\
& \\
$w$ & \multicolumn{2}{c}{$ 0.397 ^{+ 0.109 }_{- 0.097 }$} & \multicolumn{2}{c}{$ 0.453 ^{+ 0.182 }_{- 0.148 }$} & \multicolumn{2}{c}{$\equiv 0$} & \multicolumn{2}{c}{$ 0.594 ^{+ 0.123 }_{- 0.117 }$} \\ 
%%%%%%%%
& \\
$\sigma_{\rm int}$ & $\equiv 0$ & $\equiv 0$ & $<0.055$ & $<0.073$ & \multicolumn{2}{c}{$0.054^{+0.025}_{-0.028}$} & $\equiv 0$ & $\equiv 0$\\
%%%%%
 & \\
 \hline \\
  $\langle E(B-V)\rangle$ & $0.052^{+0.029}_{-0.030}$ & $0.102^{+0.036}_{-0.033}$ & $0.058^{+0.039}_{-0.037}$ & $0.093^{+0.037}_{-0.034}$ & 
  \multicolumn{2}{c}{$0.103^{+0.042}_{-0.039}$} &
  $0.120^{+0.036}_{-0.036}$ & $0.101^{+0.035}_{-0.035}$ \\
  & \\
 $\Delta$BIC & \multicolumn{2}{c}{$0$} & \multicolumn{2}{c}{+10.8} & \multicolumn{2}{c}{$+6.4$} & \multicolumn{2}{c}{$-$} \\
\hline

\end{tabular}
\caption{Best fit hyperparameters of prior probability distributions describing latent variables in two distinct supernova populations 
(see Table~\ref{hyperparameters} for the notation). The constraints were obtained for the baseline model ("baseline"), its extension 
including intrinsic scatter in both supernovae populations as extra free parameters ("baseline+$\sigma_{\rm int}$") and the baseline 
model assuming a single population ("single population"). The table shows the results from fitting all three models to \textit{SuperCal} supernova sample and independently the baseline model to the Foundation DR1 sample ("Foundation"). Best fit 
results are provided as the posterior mean values and errors of the credibility range containing 68 per cent of the marginalised 
probabilities (except for $\sigma_{\rm int}$ in the "baseline+$\sigma_{\rm int}$" model where we show 68 per cent upper limits). Hyperparameters 
which are not shared by the supernova population are shown in designated columns of each model. Shared parameters, e.g. 
$\{\alpha,\beta,\widehat{R_{\rm B}},\sigma_{R_{\rm B}},\gamma\}$ in the baseline mode, are shown once between the columns 
corresponding to the supernova populations. The two bottom lines show the mean colour excess $\langle E(B-V)\rangle$ 
derived from the Markov chains and the Bayesian Information Criterion $BIC=k\ln(N)-2\ln L_{\rm max}$, where $N$ is the number of supernovae, 
$k$ is the number of model parameters and $L_{\rm max}$ is the maximum likelihood.
}
\label{bestmodels}
\end{center}
\end{table*}

The two populations appear to have quite different distributions of intrinsic colours and reddening. The observed supernova colours 
in population 1 are primarily driven by intrinsic colour and to a much lesser extent by dust reddening. This trend is reversed in population 2
supernovae whose observed colours are clearly driven by reddening due to dust. The relatively narrow ranges of reddening in population 1 and 
intrinsic colours in population 2 reduce the constraining power for $\{\widehat{R_{B}},\sigma_{R_{B}},\gamma\}$ in population 1 and $\beta$ in 
population 2. These poorly constrained parameters are however consistent with their analogs from the opposite population and are well within the intervals 
given by their uncertainty. This shows that keeping $\{\widehat{R_{B}},\sigma_{R_{B}},\gamma,\beta\}$ independent in the two 
supernova populations cannot be constrained by the data. For this reason we will hereafter assume that these parameters are shared by both 
supernova populations, although both populations contribute unevenly to the final constraints. An analogous reduction in the parametrisation 
is justified in the case of the $\alpha$ parameter. Despite a striking difference between the distributions of $x_{1}$ in both populations, the 
linear relations between $m_{\rm B}$ and $x_{1}$ are consistent with having the same slopes in both populations. Assuming identically 
equal $\{\widehat{R_{B}},\sigma_{R_{B}},\gamma,\alpha,\beta\}$ parameters in both supernova populations defines the main class of models 
which we explore in more detail in our study.

The two left columns in Table~\ref{bestmodels} show constraints on all free parameters of our model with or without intrinsic scatter in 
both populations included as free parameters. Comparing these two cases we can see that including intrinsic scatter has virtually no impact 
on the remaining parameters. Furthermore, the best fit model is clearly consistent with vanishing intrinsic scatter in both populations. 
We find that that the maximum likelihood $\sigma_{\rm int}$ is $\{0.03,0.00\}$ and the corresponding $1\sigma$ upper limits are $\{0.06,0.07\}$. 
The model with the intrinsic scatter is also strongly disfavoured with respect to its version with $\sigma_{\rm int}\equiv 0$ in terms of the Bayesian 
Information Criteria yielding $\Delta BIC=+10.7$. This is the evidence that our two-population model provides a complete description of the supernova 
data on the Hubble diagram without invoking any extra scatter. In what follows, we shall consider its fully optimised version in which intrinsic scatter 
vanishes in both supernova populations. We will hereafter refer to this model as the baseline model.

Figure~\ref{baseline} and Table~\ref{bestmodels} show constraints on parameters obtained for the baseline model. As mentioned above, the primary 
difference between the two populations isolated by the model lies in their stretch parameter distributions. The two population have comparable 
ratios: $40$ per cent for population 1 (fast declining supernovae) and $60$ per cent for population 2 (slowly declining supernovae). Supernovae 
in population 2 appear to be intrinsically bluer than their analogs from population 1 
and their intrinsic colours are less scattered around the distribution peak. These supernovae are also more affected by dust reddening and extinction. 
The distribution of apparent colours in this population results primarily from dust 
reddening and a typical $E(B-V)$ colour excess is 2 times larger than in population 1. We also find a tentative trend for population 1 supernovae being 
intrinsically brighter than their analogs from population 2 conditioned to have the same stretch parameter and intrinsic colour. 

\subsection{Prior distribution of $E(B-V)$}

\begin{figure*}
	\centering
	\includegraphics[width=\linewidth]{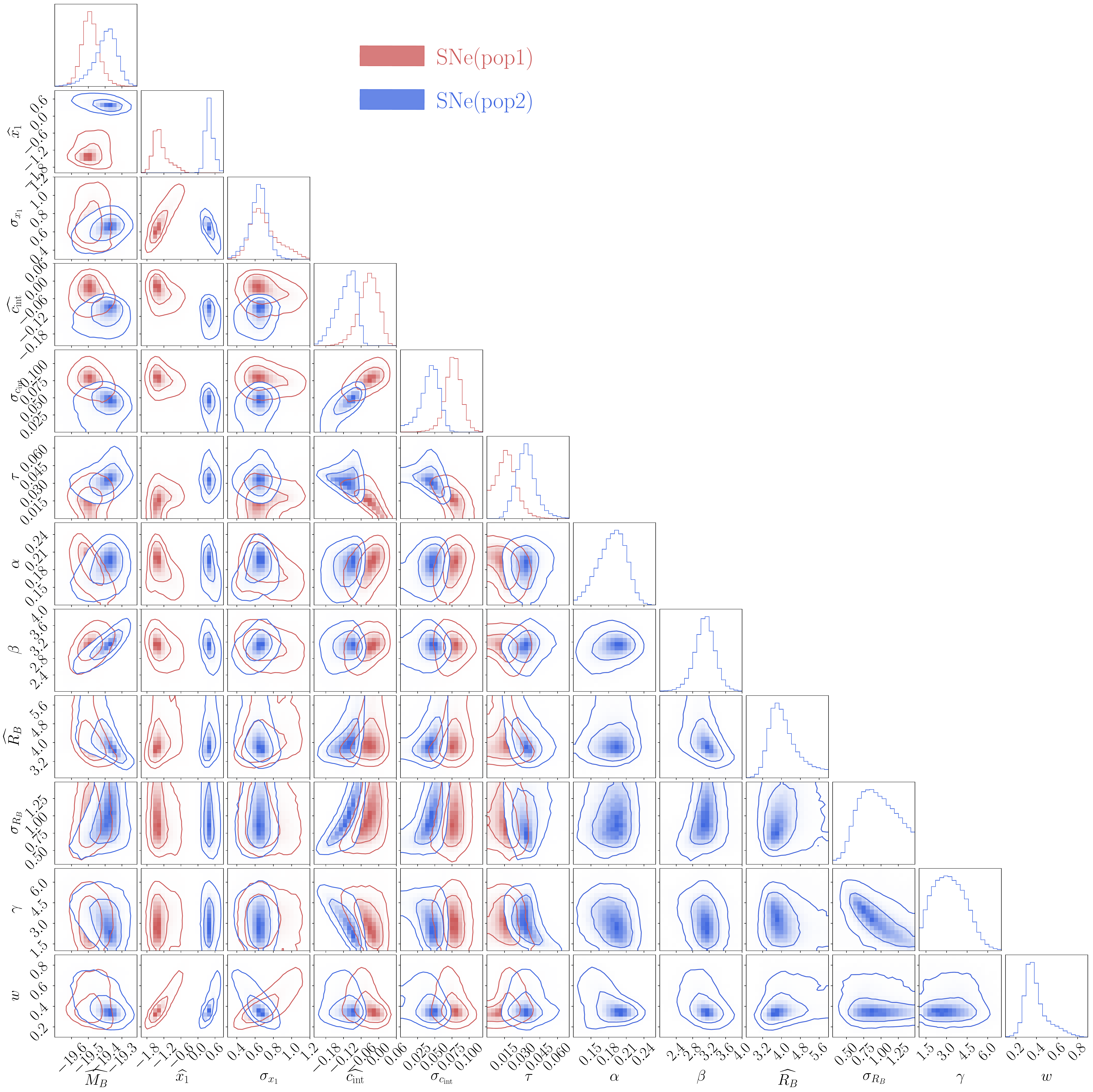}
	\caption{Constraints on hyperparameters of the baseline model obtained for the \textit{SuperCal} supernova sample. The red and blue contours show constraints 
	on parameters in the two supernova populations: population 1 (fast declining supernovae, red) and population 2 (slowly declining supernovae, blue). 
	For the sake of better readability, the corner plot is compressed by overlaying the panels with the corresponding sets of hyperparameters. The contours 
	show $1\sigma$ and $2\sigma$ confidence regions containing 
	68 and 95 per cent of 2-dimensional marginalised probability distributions.
	}
	\label{baseline}
\end{figure*}

A very important finding 
of our analysis is a clear preference for a peaked (two-tailed) distribution of $E(B-V)$ colour excess ($\gamma>1$) over the exponential 
distribution ($\gamma=1$). This property has a direct impact on implied supernovae intrinsic colours. As shown in Figure~\ref{baseline}, supernovae 
in both populations are intrinsically bluer (more negative $\widehat{c_{\rm int}}$) than in models assuming an exponential distribution of $E(B-V)$. We find evidence for a two-tailed distribution 
of $E(B-V)$ in analysis of both supernova samples (\textit{SuperCal} and Foundation DR1). The property is not related to two-population assumption and it holds for reduced models assuming a single supernova population (see Table~\ref{bestmodels}).

\subsection{Comparison of two supernova samples}

It is instructive to compare the properties of the baseline model inferred from the \textit{SuperCal} supernova sample to those from the Foundation DR1 sample. 
Initial fits with the Foundation data are affected by  insufficient constraining power for $\beta$ with a nearly flat marginalized posterior distribution. This most 
likely results from slightly larger errors on the colour parameters $c$ ($20$ per cent larger than in the \textit{SuperCal} supernova sample) and from the fact that 
the Foundation data favour narrower distributions of intrinsic colours in both supernova populations for which measuring the slope $\beta$ becomes 
less precise. We circumvent this problem by adopting constraints on $\beta$ from the \textit{SuperCal} sample as a prior probability in the analysis 
of the Foundation data. Comparing the results obtained from the two supernova samples (see Table~\ref{bestmodels}) it is apparent that virtually 
all properties of the baseline model are quite similar. In particular, the same peaked shape of the distribution in $E(B-V)$ with $\gamma\approx 3.9$ 
is favoured over the exponential model by both supernova sample. Both data sets place fully consistent constraints on the distribution of $R_{\rm B}$ 
with $\widehat{R_{\rm B}}\approx 4$ and $\sigma_{R_{\rm B}}\approx0.9$. The difference between the two supernova populations in terms of 
dust reddening and intrinsic colours found for the \textit{SuperCal} sample are less pronounced in the Foundation sample. This is partially caused 
by a weaker constraining power of the Foundation data resulting in larger errors of hyperparameters related to the observed supernova 
colour, e.g. $\sigma_{\rm c_{\rm int}}$ and $\tau$. The Foundation sample seems to favour a slightly elevated fraction of population 1 supernovae, 
although the difference is not statistically significant ($1.3\sigma$). 

The Foundation sample appears to have less colour-dependent scatter than the \textit{SuperCal} sample. When refitting the baseline model with 
free intrinsic scatter in both supernova populations, we find an upper limit for scatter in the extinction parameter, i.e. $\sigma_{R_{\rm B}}<0.34$, 
and a clear preference for achromatic scatter with $\sigma_{\rm int}=0.1\pm0.02$. Therefore, the scatter $\sigma_{R_{\rm B}}$ obtained for the Foundation sample and shown in Table~\ref{baseline} is to some extent driven by the assumption that intrinsic scatter vanishes in the baseline model.

\subsection{Robustness tests}

Repeating our fits using supernovae with the most recent updates of light curve parameters and redshifts from the \textit{Pantheon+} compilation, we 
find a broad agreement with the results obtained for the \textit{SuperCal} sample. In Appendix B we show constraints on parameters of the baseline model 
with the mean extinction parameter $\widehat{R_{\rm B}}$ allowed to be independent in the two supernova populations. The properties of the best fit model resemble 
closely those obtained for the \textit{SuperCal} and most parameters derived from the two data sets remain well within their errors. The only noticeable difference 
is a stronger degeneracy between intrinsic colour, reddening and $\widehat{R_{\rm B}}$ in population 1 (see Fig.~\ref{baseline_pan}).

In order to test a potential impact of survey limiting magnitudes on our results, we repeat our analysis including redshift-dependent bias estimated for 
low-$z$ surveys in \citet{Scolnic2018}. Considering two models explored in \citet{Scolnic2018}, we find that the bias has a negligible impact with relative 
shifts of the best fit values not exceeding 20 per cent of the uncertainties. Furthermore, we also checked that restricting our analysis to nearby 
supernovae at $z<0.08$, which effectively mitigates potential observational biases in a model independent way, returns best fit parameters well within 
the errors of the best fit model based on the whole sample.

\subsection{Hubble residuals}

The baseline model provides a complete probabilistic description of type Ia supernova as standardisable candles. The unavoidable intrinsic scatter 
on supernova Hubble diagrams constructed by applying the standard Tripp formula (or its extensions, such as including additional corrections related to the host stellar 
mass) is fully accounted for in terms of dust extinction and differences between intrinsic and extrinsic properties of two supernova populations. 
In order to appreciate the explanatory potential of our model, we compute the distribution of supernovae in the space of light curve observables 
$\{m_{\rm B}-\mu,x_{1},c\}$ predicted by the baseline model and compare to the actual distribution of supernovae from the observational sample. 
We calculate the predicted distribution in a Monte Carlo way by sampling from the prior distributions of all latent variables and generating marginalised distributions in the 
observable space based on eq.~(\ref{margin_prior}). We use best fit parameters of the baseline model ($\sigma_{\rm int}\equiv 0$) measured from 
the \textit{SuperCal} supernova sample. We subtract the effect of distances by considering $m_{\rm B}-\mu$ instead of $m_{\rm B}$.

\begin{figure*}
     \centering
     \begin{subfigure}[t]{0.49\textwidth}
         \centering
         \includegraphics[width=\textwidth]{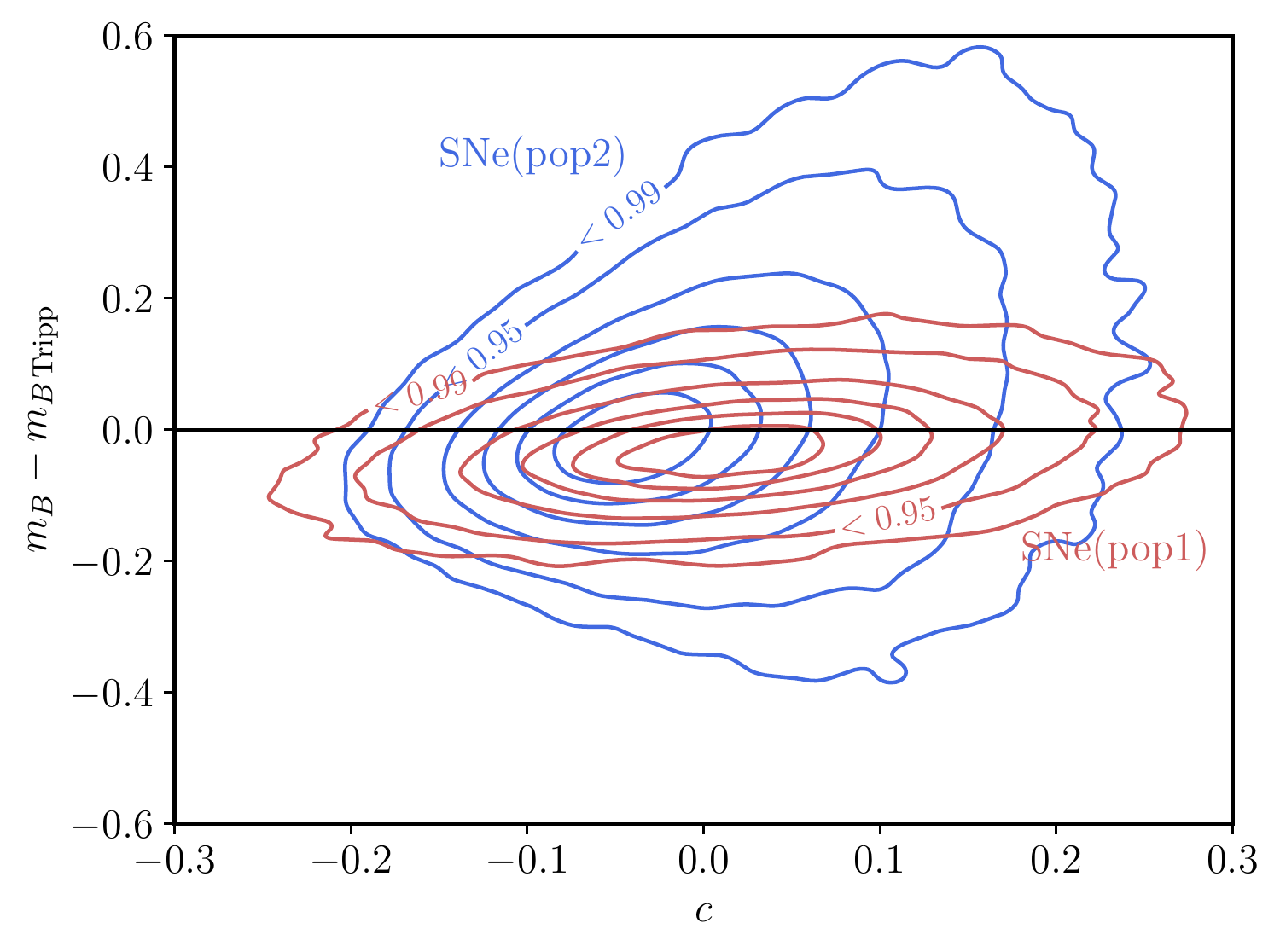}
         \caption{}
         \label{fig:y equals x}
     \end{subfigure}
     \hfill
     \begin{subfigure}[t]{0.49\textwidth}
         \centering
         \includegraphics[width=\textwidth]{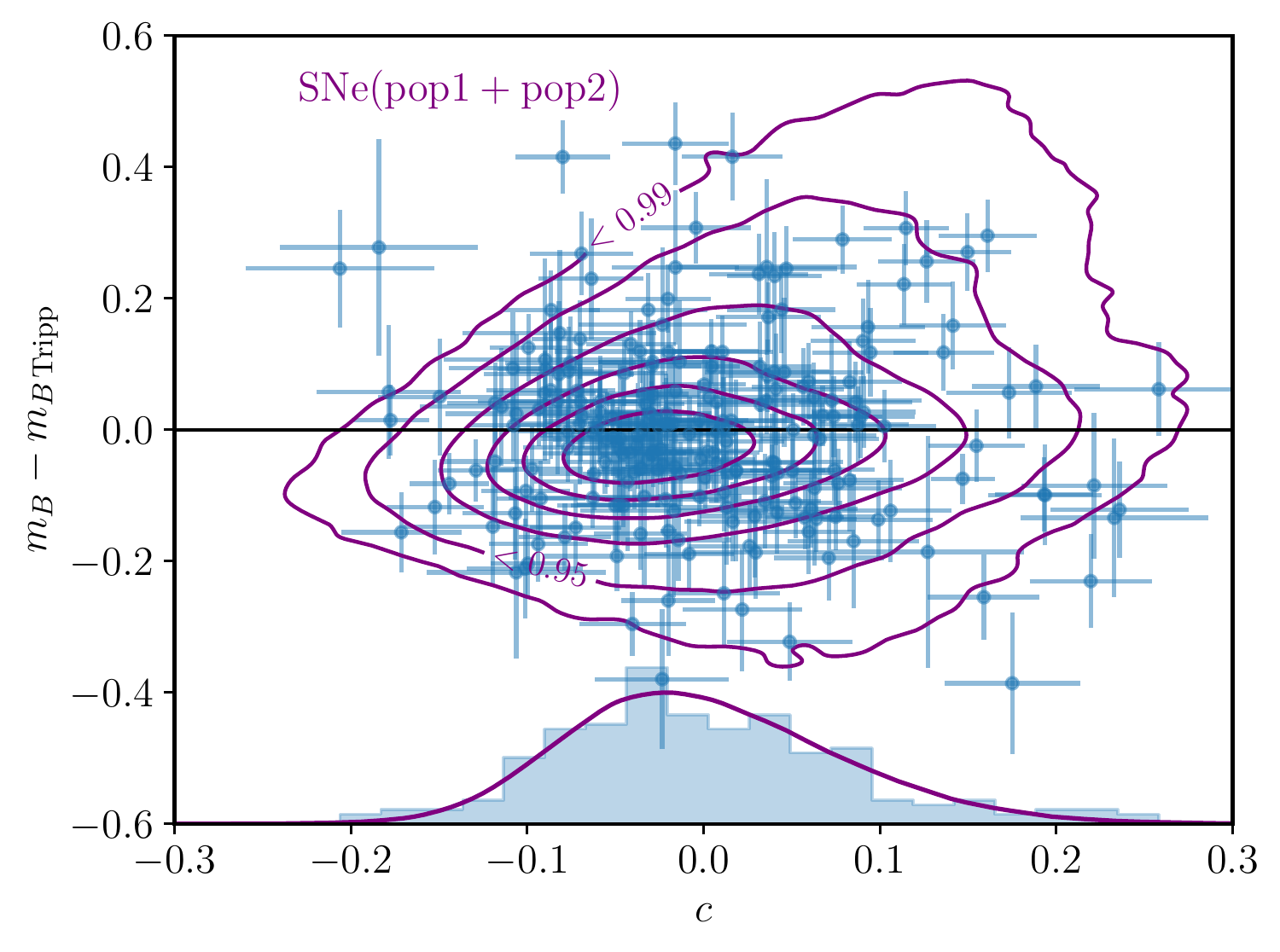}
         \caption{}
         \label{fig:three sin x}
     \end{subfigure}
     \centering
     \begin{subfigure}[b]{0.49\textwidth}
         \centering
         \includegraphics[width=\textwidth]{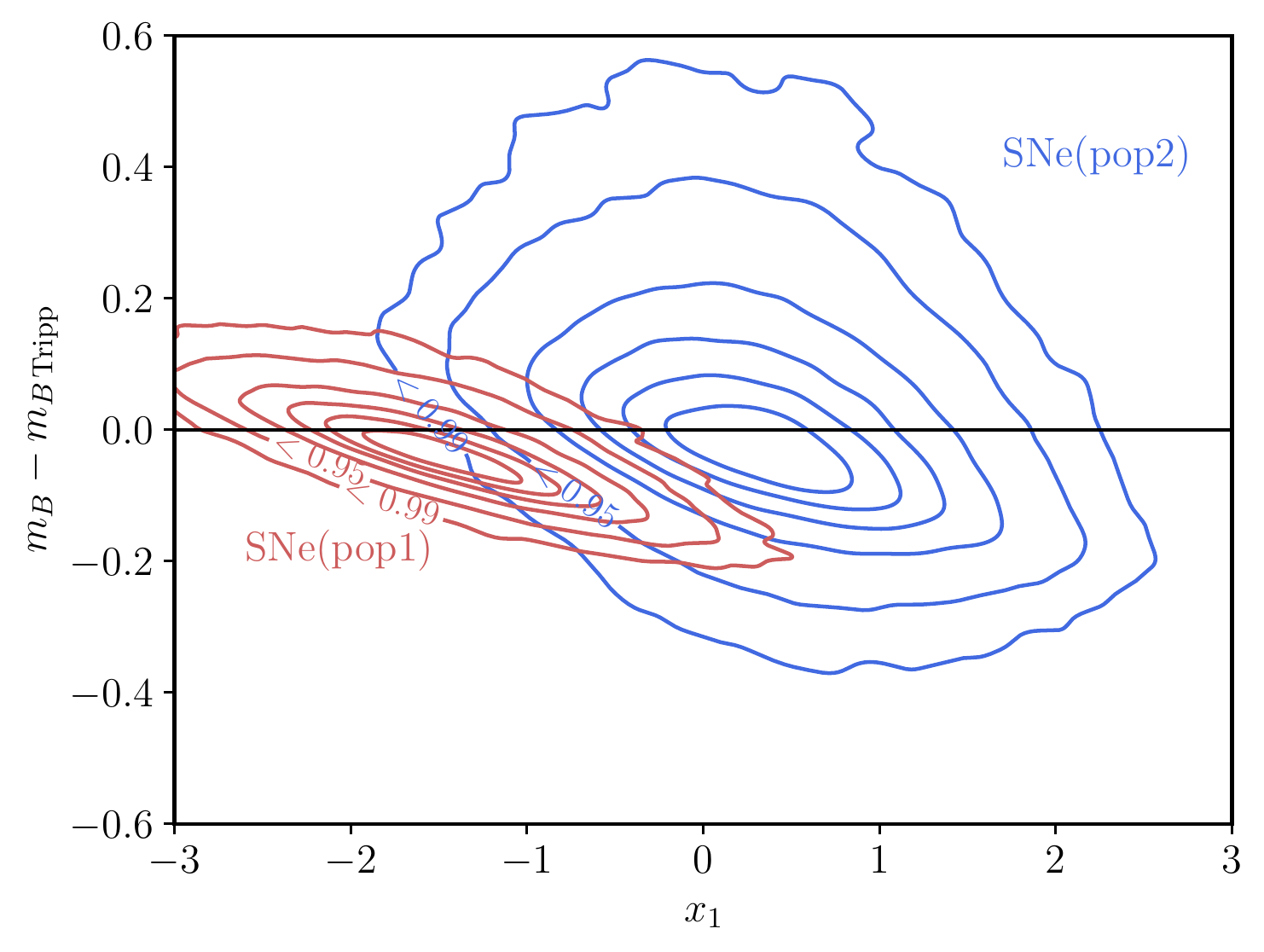}
         \caption{}
         \label{fig:y equals x}
     \end{subfigure}
     \hfill
     \begin{subfigure}[b]{0.49\textwidth}
         \centering
         \includegraphics[width=\textwidth]{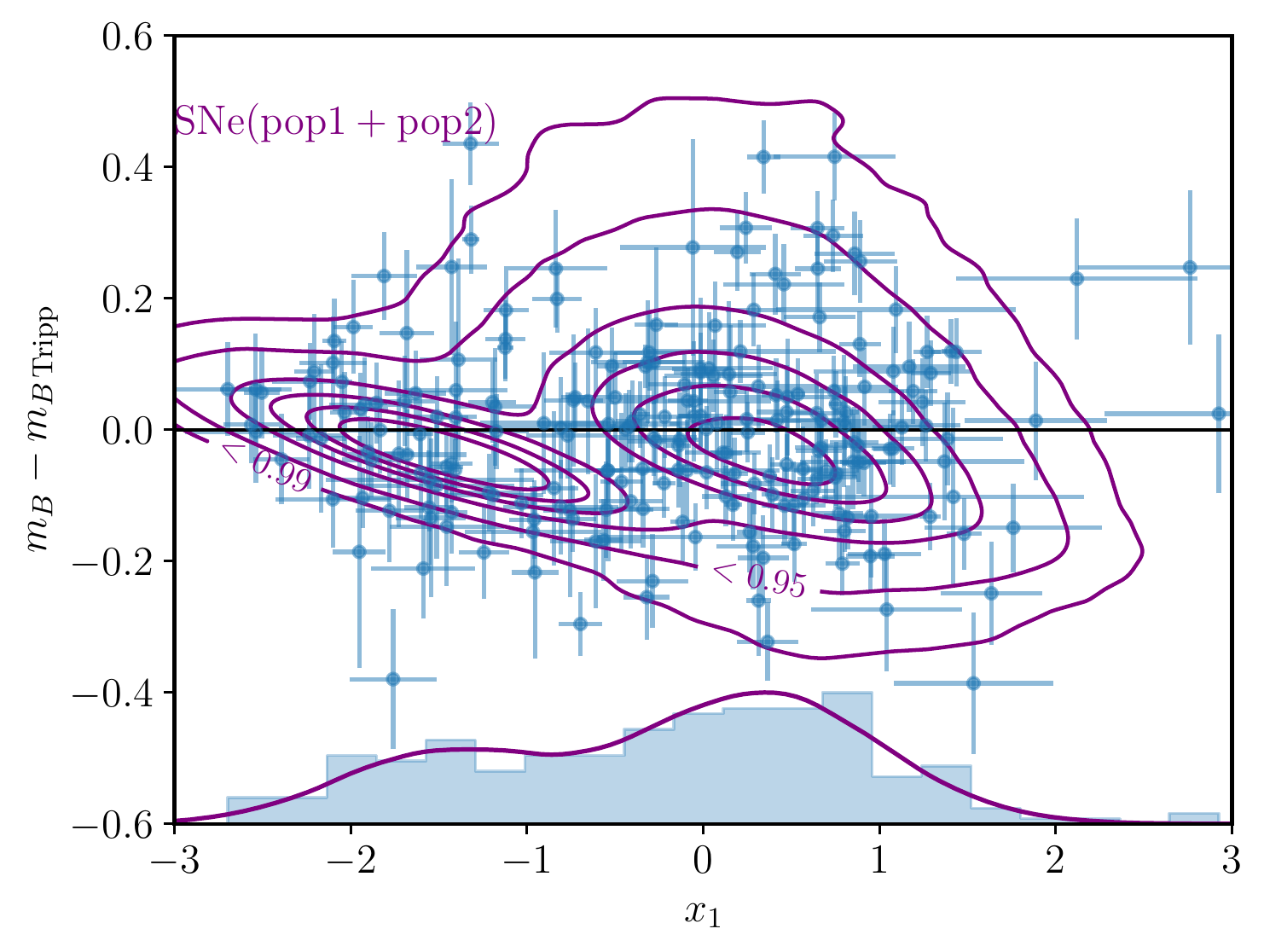}
         \caption{}
         \label{fig:three sin x}
     \end{subfigure}
        \caption{Distributions of the Hubble residuals in type Ia supernova peak magnitudes 
        (standardised using baseline model relative to Tripp calibration), as a function of colour parameter $c$ or stretch parameter $x_{1}$. The symbols show supernova 
        data from the \textit{SuperCal} supernova sample used in this study, while contours show the distributions implied by the 
        best fit baseline model. The left panels illustrate the effect of differences between the distributions 
        of latent variables in the two supernova populations (population 1 -- fast declining; population 2 -- slow declining) 
        on the observed light curve parameters. The right panels demonstrate that the best fit baseline model provides a complete and accurate description 
        of the apparent Hubble residuals including (i) bimodality in the distribution of $x_{1}$, (ii) asymmetries 
        between positive and negative residuals and (iii) scatter in the Hubble residuals increasing with the 
        supernova colour parameter. The histograms show the distributions of observed supernova light curve parameters. They are compared to the best fit model shown 
        with the corresponding solid lines. All apparent features in the distribution of the Hubble residuals and observed supernova colours 
        are accounted for in terms of extinction properties and differences between intrinsic (intrinsic colour) and extrinsic 
        (dust reddening) properties of the two model populations.
        }
        \label{residuals}
\end{figure*}

It is instructive to express supernova magnitude $m_{\rm B}$ relative to a model given the Tripp formula \citep{Tripp1998}, i.e.
\begin{equation}
m_{\rm B\,Tripp}=M_{\rm B\,T}+\mu(z)-\alpha_{\rm T} x_{1}+\beta_{\rm T} c.
\end{equation}
This conversion reduces to a change of one variable in the model's predictions, but for the actual supernova data it generates genuine residuals expected 
on the corresponding Hubble diagram. For the sake of complete consistency with the data we use $\{M_{\rm B\,T},\alpha_{\rm T},\beta_{\rm T}\}$ measured directly 
from the supernova sample. Performing a fit based on the likelihood given by
\begin{align}
\ln L\propto & -\frac{1}{2}\sum_{i}^{N}\frac{[m_{\rm B\,i}-m_{\rm B\,Tripp}(z_{i},c_{i},x_{1\,i})]^{2}}{\sigma_{\rm tot\,i}^{2}+\sigma_{\rm int\,T}^{2}} \\
                  & -\frac{1}{2}\sum_{i}^{N}\ln(\sigma_{\rm tot\,i}^{2}+\sigma_{\rm int\,T}^{2})
\end{align}
where $\sigma_{\rm tot\,i}$ is the total uncertainty including contributions from all elements of the covariance matrix and uncertainty related 
to peculiar velocities \citep[see e.g.][]{Brout2021,Wojtak2022}, and $\sigma_{\rm int\,T}$ is an extra free parameter describing intrinsic scatter, 
we obtain $M_{\rm B\,T}=-19.338\pm0.009$, $\alpha_{\rm T}=0.128\pm0.008$, $\beta_{\rm T}=3.00\pm0.12$ and $\sigma_{\rm int\,T}=0.11\pm0.008$ 
(assuming consistently the same cosmological model with $H_{0}=70\,{\rm km}\,{\rm s}^{-1}{\rm Mpc}^{-1}$ as in the main analysis).

Figure~\ref{residuals} compares the distribution of residuals in Tripp formula
supernova magnitudes on the Hubble diagrams relative to the baseline model's predictions,
as a function of stretch parameter $x_{1}$ or colour parameter $c$.
The panels demonstrate how precisely the model 
captures all subtle features in the distribution of the Hubble residuals and supernova light curve parameters. In particular, the model reproduces the conspicuous
bimodality in the stretch parameter and a clear trend of Hubble residuals increasing with the colour parameter. The latter is primarily 
a manifestation of a rather large scatter in $R_{\rm B}$, i.e. $\sigma_{\rm R_{\rm B}}\approx0.9$, giving a characteristic pattern of divergent isodensity 
contours towards positive colour parameters ($c>0$) in population 2 supernovae (whose observed colours are dominated by dust reddening). Here we 
can also see excess of positive Hubble residuals at $0\lesssim c \lesssim 0.2$ (about 2 times more supernovae with residuals $(m_{\rm B}-m_{\rm B\,Tripp})>0.2$ than those with $(m_{\rm B}-m_{\rm B\,Tripp})<-0.2$) which can be ascribed to the fact that the bulk of $R_{\rm B}$ 
values allowed by the model (including $\widehat{R_{\rm B}}$) are larger than the colour correction coefficient $\beta_{\rm T}$ of the Tripp calibration. 
Both the apparent asymmetry and colour-dependent scatter in the Hubble residuals have been described in several studies \citep[see e.g.][]{Brout2021,Popovic2021}.
These are anomalous features in the standard framework based on the Tripp calibration. They play a key role in disentangling the effects 
of intrinsic colours from dust reddening in both supernova populations.

Comparing Hubble residuals in supernova populations from our modelling and analogous populations of young and old supernovae from \citet{Rigault2020}, 
we find a similar scatter in population 2 ($0.13$) as in the corresponding young supernovae, but two times smaller scatter in population 1 than in the corresponding 
old supernovae. To a large extent the apparent difference results from the fact that the old population of \citet{Rigault2020} does not consists solely of fast 
decliners, but mixes high- and low-stretch supernovae.

\section{Discussion}

The baseline model developed in this study is the first Bayesian hierarchical model of type Ia supernovae which provides 
a compete explanation of Hubble residuals in terms of intrinsic properties of type Ia supernovae and extrinsic effects of dust. 
The key new element is the theoretically and observationally motivated premise that type Ia supernovae originate from two distinct populations 
most likely associated with star-forming and passive environments. The model recovers several important findings obtained with 
previous Bayesian analyses assuming a single supernova population, but it also refines many details which are relevant 
to building a consistent holistic picture of type Ia supernovae and their environments. In the following, we describe the physical 
properties implied by our model and put them a broader context of type Ia supernova studies.

\subsection{Two populations}

A two-population model is strongly favoured by the supernova data. Fitting its analog with a single population we find that 
the model can account for only a fraction of the intrinsic scatter in the Hubble residuals (see Table~\ref{bestmodels}). 
In a single-population model, a physically motivated dust model with a wide range of $R_{\rm B}$ values reduces the intrinsic scatter from 
$\sigma_{\rm int}\approx 0.11$ for the Tripp calibration to $\sigma_{\rm int}\approx0.06$. This effect is 
also found in several independent studies employing similar single-population Bayesian hierarchical models \citep{Mandel2017,Thorp2021,Mandel2022}. Further reduction of the intrinsic scatter, however, can only be achieved by considering two supernova populations. In term of the Bayesian Information Criterion, we find the the baseline model is strongly favoured by the \textit{SuperCal} supernova sample over its single-population analog with $\Delta BIC=-6.8$.

The two supernova populations emerging from our baseline model can be tentatively interpreted as 
supernovae originating from an old (delayed) progenitor channel (population 1) and a young (prompt) 
progenitor channel (population 2). Assuming that there exists a rather unambiguous mapping between the model populations and these progenitor channels, 
we expect that the weight parameter $(1-w)$ in our model can be regarded as 
an estimate of the fraction of supernovae from the prompt progenitor channel. The most rigorous 
measurement of the population fraction was obtained from fitting redshift dependance of type Ia supernova 
volumetric rate \citep{Rodney2014}. The key idea of this measurement is that the observed volumetric rate 
can be probabilistically split into two distinct components associated with the prompt progenitor channels 
(following the star formation history) and the delayed progenitor channel (lagged behind the star formation 
history with a characteristic time delay distribution $\propto t^{-1}$ expected for double-degenerate 
binary systems) \citep{Maoz2014}. Analysis of type Ia supernova volumetric rate measurements as a function of redshift showed 
that the global fraction of prompt type Ia supernovae is $f_{\rm p}=0.53^{+0.14}_{-0.36}$ \citep{Rodney2014}. 
Taking into account the star formation history and the time delay distribution assumed in \citet{Rodney2014}, 
this implies that the prompt channel fraction at redshift $z=0$ is $f_{\rm p}(z=0)=0.30^{+0.08}_{-0.20}$. Independent estimates of $f_{\rm p}$ can be derived from observations of blueshifted Na~I~D absorption features in supernova spectra. These absorption features are interpreted as a signature of circumstellar material from a non-degenerate companion star of supernova progenitor and thus they can be used to differentiate between single and double degenerate progenitor channels (or related delayed and prompt channels). Observations yield $f_{\rm p}(z=0)$ ranging between $0.2$ \citep{Maguire2013} to $0.5$ \citep{Phillips2013}. The corresponding fractions measured in our analysis, i.e. $1-w=0.60\pm{0.1}$ from the main and $1-w=0.41\pm{0.12}$ from the Foundation DR1 sample, appear to be consistent with the upper limit of the above-mentioned estimates. \citet{Andersen2018} modelled the SN Ia rate as a function of specific star formation rate of their host galaxies and found $f_{\rm p}(z=0)=0.10^{+0.10}_{-0.06}$.

Population 2 (slowly declining) supernovae appear to be on average intrinsically bluer than population 1, with $\Delta\widehat{c_{\rm int}}=-0.077\pm0.032$. Their mean intrinsic colour is $\Delta c_{\rm int}\approx-0.03$ bluer than the previous estimates from analyses assuming single supernova population \citep{Popovic2021,Brout2021}. Slowly declining supernovae (population 2) also seem to be $0.09\pm0.07$~mag fainter. This is only an indicative offset, but it is consistently found in both supernova samples. We note that a similar trend was found for the populations of fast and slowly declining supernovae selected by the stellar age supernova environments \citep{Rigault2013,Rigault2020,Maoz2014}.

\subsection{Dust}

Supernovae from the two populations discerned by the baseline model exhibit different degrees of dust reddening 
as measured by the scale parameter $\tau$. Based on the model's parameters we conclude that population 1 (fast declining) 
supernovae are associated with dust-poorer environments with the mean colour excess $\langle E(B-V)\rangle\approx 0.052\pm0.030$, 
while population 2 (slowly declining) supernovae with dust-rich environments with $\langle E(B-V)\rangle\approx 0.102\pm0.035$. This difference is 
not surprising given a wide range of observations showing that type Ia supernovae similar to those in population 2 
(slowly declining light curves) are typically found in highly star-forming and thus dust rich environments 
\citep[see e.g.][]{Rigault2020}. It is clearly supported by the data of the \textit{SuperCal} supernova sample for which 
we find a significance level of $2.3\sigma$ (higher than expected from a summary statistics shown in Table~\ref{bestmodels} 
due to a strong correlation between $\tau$ parameters in the two populations). It is not apparent in the Foundation sample, 
most likely due to a weaker constraining power of the Foundation data.

\begin{figure}
	\centering
	\includegraphics[width=\linewidth]{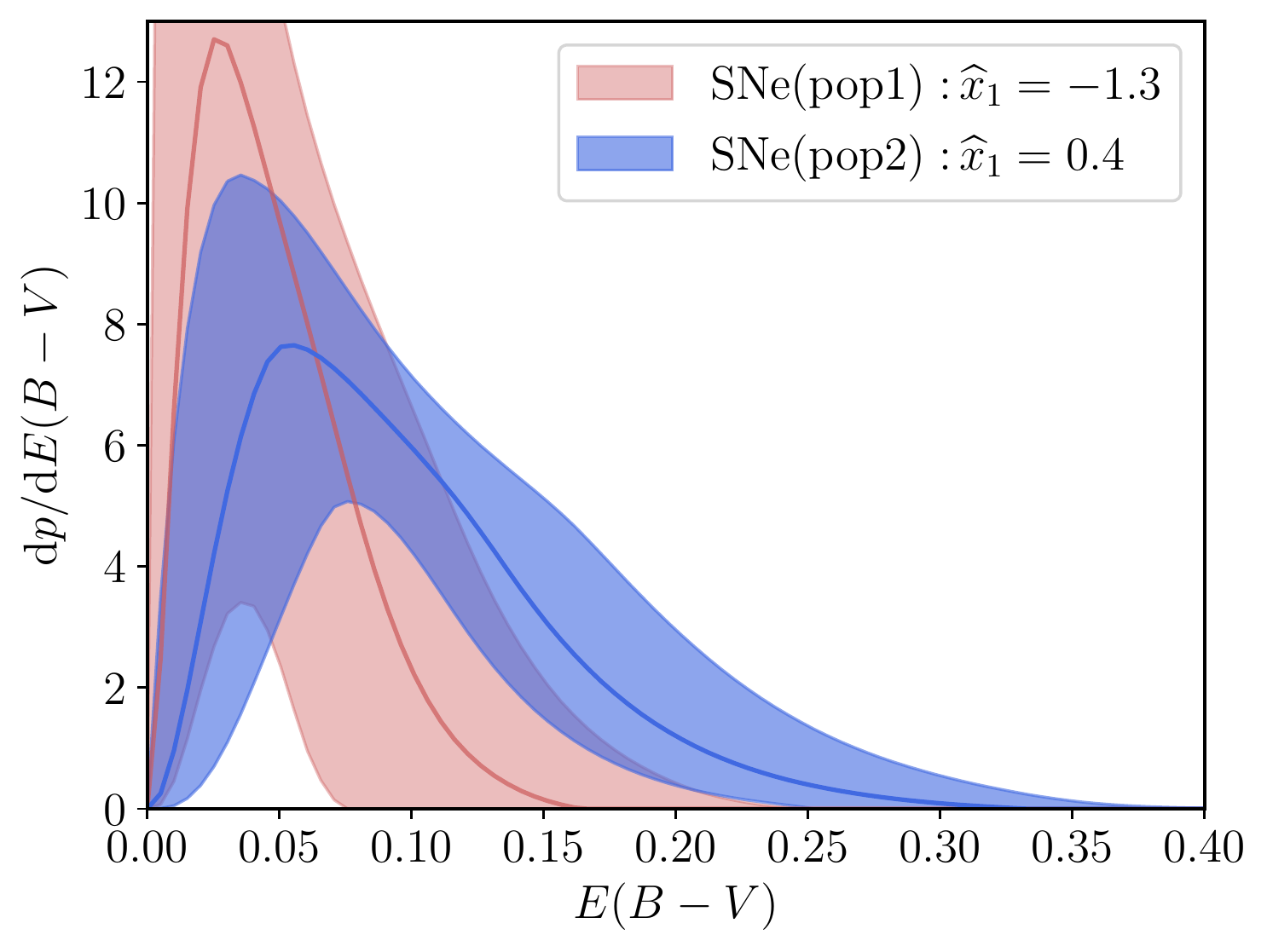}
	\caption{Distribution of colour excess $E(B-V)$ in the two supernova populations emerging from the baseline model 
	fitted to the \textit{SuperCal} supernova sample. The lines and contours show the best fit model predictions and the 
	corresponding $68$ per cent uncertainties computed by sampling from the corresponding MCMC. Both 
	populations are described by peaked distributions with a maximum at $E(B-V)>0$. The scale of reddening 
	in population 2 (slowly declining) supernovae is $\sim 2$ times larger than in population 1 (fast declining supernovae).
	}
	\label{reddening}
\end{figure}

Dust extinction inferred from both supernova samples is described by a wide range of extinction coefficients. The mean value of $R_{\rm B}$, i.e. $\widehat{R_{\rm B}}\approx 4.1$, is fully consistent with an average extinction measured directly in the Milky Way \citep{Fitzpatrick1999,Schlafly2016} and nearby galaxies \citep{Draine2003}. The range of extinction coefficients in individual sight lines allowed by the scatter $\sigma_{\rm R_{\rm B}}=0.9$ overlaps largely with a range $3\lesssim R_{\rm B}\lesssim 7$ measured in the Milky Way \citep{Fitzpatrick2007,Draine2003}. We also note that the recently found anomalously high colour correction of type Ia supernovae in host galaxies with observed Cepheids \citep[$\beta_{\rm T}=4.6\pm0.4$;][]{Wojtak2022} is consistent with a typical extinction coefficient found in our analysis. We interpret this coincidence as an indication that supernovae in the calibration sample of the SH0ES program originate primarily from population 2 (supernovae with colours driven predominantly by dust reddening). In the recent SH0ES 
measurement of the Hubble constant, the adopted approach to mitigating differences between supernovae in Cepheid host galaxies and the Hubble flow relies on selecting supernovae in late-type galaxies similar to those in the calibration sample \citep{Riess2022}.

Dust extinction is primarily constrained by population 2 supernovae whose observed colours are dominated by dust reddening. Constraining  $\widehat{R_{\rm B}}$ independently in population 1 by fitting a one-parameter extension to the 
baseline model to the \textit{SuperCal} supernova sample yields $\widehat{R_{\rm B}}=3.6^{+1.3}_{-1.2}$ and shows no indication of any 
difference between extinction properties in the two supernova populations. Our constraints on the distribution of $R_{\rm B}$ are also broadly consistent with 
the previous estimates based on Bayesian hierarchical models with a single supernova population \citep[see e.g.][]{Thorp2021} 
and alternative estimates based on modelling supernova colours in multi-band observations, for a similar range of reddening with $E(B-V)\lesssim0.3$ \citep{Burns2014}. 
As demonstrated in Table~\ref{bestmodels}, alternating between single- and two-population models has a rather negligible 
impact on the estimation of the mean and scatter of the extinction coefficient $R_{\rm B}$.

Constraints on the shape parameter $\gamma$ demonstrate that the supernova data favour a peaked (two-tailed) distribution of 
colour excess $E(B-V)$, well represented by a gamma distribution with $\gamma\approx3.3$, over the commonly assumed 
exponential model \citep[][]{Thorp2021,Mandel2017,Brout2021}. The shape of the distribution as well as the 
difference between the two supernova populations in terms 
of the mean colour colour excess are illustrated by Fig.~\ref{reddening}. The distributions shown in the figure were computed
for the baseline model fitted to the \textit{SuperCal} supernova sample. Compared to its exponential analog, peaked distributions of colour 
excess $E(B-V)$ enhance the effect of dust reddening in the observed supernova colours resulting in intrinsically bluer 
implied intrinsic colours. This property is shown in Figure~\ref{baseline} as a correlation between $\widehat{c_{\rm int}}$ and $\gamma$.

The role of supernova populations in explaining the Hubble residuals in terms of extinction properties was realised 
to some extent by \citet{Brout2021}. They showed that deriving $R_{\rm B}$ distributions in two separate bins of supernova 
host stellar masses gives a more complete description of the Hubble residuals. Their analysis based on a single-population model for 
intrinsic supernova properties implies that high stellar mass ($>10^{10}M_{\odot}$) host galaxies exhibit substantially lower 
extinction coefficients with $\widehat{R_{\rm B}}\approx 2.5$ compared to $\widehat{R_{\rm B}}\approx 3.8$ for low-mass hosts ($<10^{10}M_{\odot}$). It is not obvious, however, which
physical processes can effectively fine tune dust properties in a way that they 
can match the apparent difference in $R_{\rm B}$ between high and low stellar mass host galaxies and explain the 
corresponding transition scale in the stellar mass. In contrast, our two-population model implies dust properties which 
can be fairly easily reconciled with existing observations without any necessary fine tuning
or need for non-standard dust properties. While extinction 
properties in all supernova host galaxies resemble closely those known from the Milky Way, the two supernova 
populations are characterised by different average column densities of dust which can be naturally associated 
with the observed differences between star formation rates of their local environments.

\subsection{Host galaxy properties}

Our model enables us to assign a probability of belonging to one of the two populations to each supernova. 
This classification 
can be performed by computing relative probabilities of finding a supernova with a given set of 
light curve parameters and originating from either of the two populations. The relative probabilities are given 
by the prior probabilities of the two supernova populations marginalised over latent variables, subject to a set 
of measured light curve parameters $\pmb{\xi_{\rm obs\,i}}$. This leads to the following probability ratio
\begin{equation}
\frac{p({\rm SN_{pop\,1}})}{p({\rm SN_{pop\,2}})}=\frac{w\int \mathcal{G}[\pmb{\xi}(\pmb{\phi});\pmb{\xi_{\rm obs\,i}},\mathsf{C_{\rm obs\,i}}]
p_{\rm prior\,1}(\pmb{\phi})\textrm{d}\pmb{\phi}}{(1-w)\int \mathcal{G}[\pmb{\xi}(\pmb{\phi});\pmb{\xi_{\rm obs\,i}},\mathsf{C_{\rm obs\,i}}]
p_{\rm prior\,2}(\pmb{\phi})\textrm{d}\pmb{\phi}}
\label{classification}
\end{equation}
which we use as a continuous variable to probabilistically classify supernovae to the model populations. We estimate 
the most likely range of the probability ratios by propagating the values of hyperparameters from the precomputed 
Markov chains to eq.~(\ref{classification}). The final estimates are then computed as the mean and scatter of the 
logarithmic probability ratios.

\begin{figure*}
     \centering
     \begin{subfigure}[t]{0.49\textwidth}
         \centering
         \includegraphics[width=\textwidth]{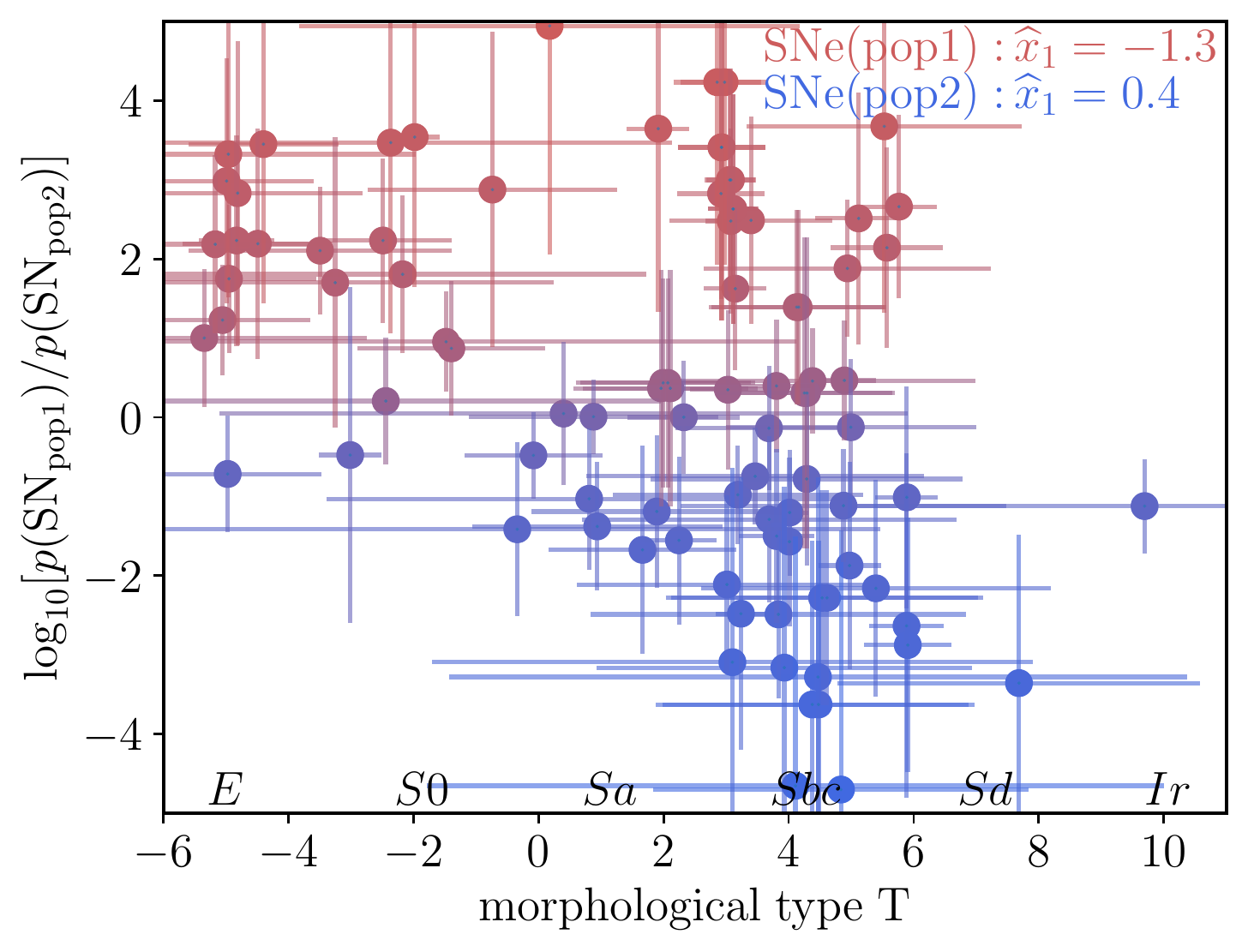}
         \caption{}
         \label{fig:y equals x}
     \end{subfigure}
     \hfill
     \begin{subfigure}[t]{0.49\textwidth}
         \centering
         \includegraphics[width=\textwidth]{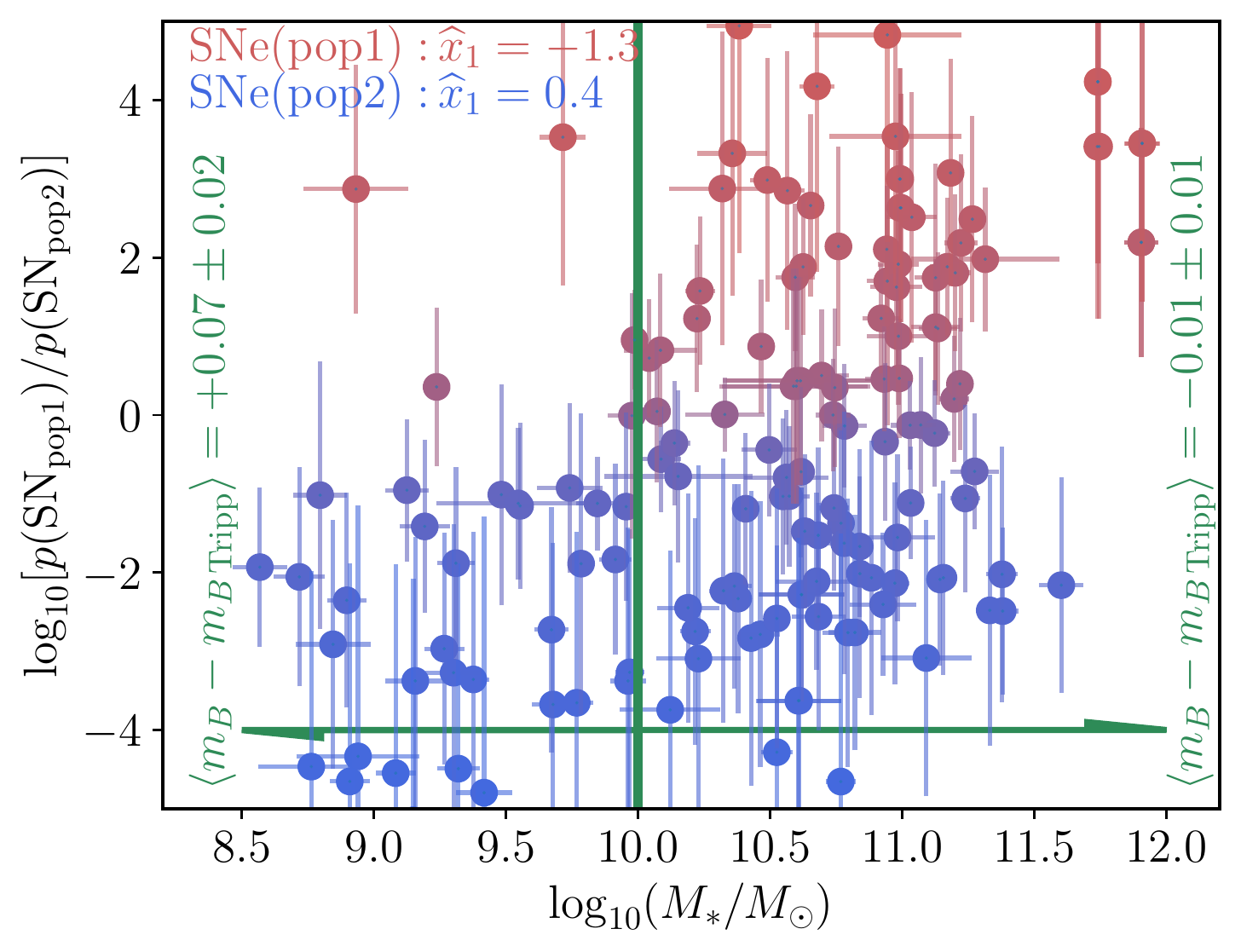}
         \caption{}
         \label{fig:three sin x}
     \end{subfigure}
        \caption{Populations of individual supernovae as a function of morphological type or stellar 
        mass of the host galaxy. The populations are assigned in a probabilistic way by comparing probabilities 
of a given supernova originating from either of the two populations in the baseline model. The assigned populations are shown both on vertical axes of the plots and as color coding of the data points. The panels show 
that population 2 (slowly declining) supernovae do not occur in early type galaxies, while population 1 (fast declining) 
supernovae are typically not found in galaxies with stellar masses below $\sim 10^{10}\rm{M_{\odot}}$. Both properties can be explained 
by assuming that population 1 (2) supernovae are associated with old (young) stellar populations and by accounting for 
how the ratio of young-to-old stellar populations changes across morphological types and stellar 
masses. The right panel also shows that the commonly used mass step correction in type Ia supernova standardisation can be attributed to
an effect of breaking the symmetry between the two supernova populations (and the associated 
stellar populations) below $\sim 10^{10}\rm{M_{\odot}}$.
        }
        \label{pop-host}
\end{figure*}

Figure~\ref{pop-host} shows the probabilistic classification of supernovae from the \textit{SuperCal} supernova sample into the two populations of the baseline model, as a function 
of the morphological type or stellar mass of the host galaxy.
The classification is based on the baseline model fitted to 
the \textit{SuperCal} supernova sample. The stellar mass estimates were obtained from modelling multi-band photometric 
observations of the host galaxies and they are are taken from the corresponding supernova catalogue 
\citep{Scolnic2015}. The morphological type is quantified by a continuous morphological variable $T$ introduced by \citet{deVa1974} and provided 
by the HyperLeda database \footnote{http://leda.univ-lyon1.fr}. This metric is estimated by combining a wide 
range of various indicators of galaxy morphology, e.g. photometric structure, colour index, or hydrogen content. 
Its values can be mapped directly into the Hubble sequence with $T=-5$ for the earliest types (E) and $T=10$ 
for the latest types (irregular galaxies), as indicated in Figure~\ref{pop-host}.

Figure~\ref{pop-host} demonstrates that the relative fractions of the supernova populations determined 
in our model vary across morphological types and stellar masses of the host galaxies. It is apparent that 
early type galaxies with $T<-1$ are primarily dominated by population 1 (fast declining) supernovae, while late type galaxies 
host both populations. This corroborates earlier studies which showed that low star formation (old stellar population) 
environments are predominantly populated by fast declining (low-stretch) supernovae \citep{Rigault2013,Rigault2020,Pruzhinskaya2020}. The same relation between the supernova populations and the underlying stellar populations can be used to explain the apparent dependency of supernova populations on the stellar mass of the host galaxy. 
Studies of galaxy properties from the Sloan Digital Sky Survey showed that galaxies with stellar masses below 
$\sim 10^{10}\rm{ M_{\odot}}$ are strongly dominated by young stellar populations, while their analogs above 
this mass limit mix young and old stellar populations \citep{Kauffmann2003}. Consequently, low-mass galaxies with stellar masses below $\sim10^{10}\rm{ M_{\odot}}$ are expected to host predominantly population 2 (slowly declining) supernovae, while high-mass galaxies are expected to produce supernovae from both populations.

The dependency on the host stellar mass shown in Figure~\ref{pop-host} is closely related to the commonly used 
mass step correction in the standardisation of type Ia supernovae \citep[see e.g.][]{Kelly2010,Scolnic2018,Smith2020}. The correction 
is intended to reduce intrinsic scatter in the supernova Hubble diagrams by subtracting the empirically measured 
difference between Hubble residuals of supernovae in galaxies with stellar masses above or below $\sim 10^{10}\rm{ M_{\odot}}$, 
where the magnitude difference is typically found in range $(0.04-0.08)$~mag ($0.08\pm0.02$ for the \textit{SuperCal} supernova sample 
used in our study). As clearly demonstrated in Figure~\ref{pop-host}, the transition mass of the mass step correction 
(see the green line) coincides with the abrupt change of relative fractions of the two supernova populations. 
This suggests that the empirical mass correction is not a fundamental relation between supernovae and 
the host galaxies, but rather an emerging property resulting from (i) the presence of two distinct supernova populations 
associated with old and young stellar populations and (ii) a step-like change of the stellar population ratio at stellar mass 
$\sim 10^{10}\rm{ M_{\odot}}$.

In order to test whether a mass step correction is needed in our model, we compute supernova Hubble residuals 
with model peak magnitudes given by eq.~(\ref{observables}). We use mean intrinsic colour and reddening $E(B-V)$ 
derived from sampling the underlying prior distribution conditioned to the measured colour parameters of each supernova. 
Predicted peak magnitudes of each supernova are then calculated using population types shown in Figure~\ref{pop-host}. 
Assuming that $R_{\rm B}$ is given by the same best fit mean value 
across all host masses, we find that our model decreases the mass step by 30 per cent from 0.08 to 0.055. Further reduction 
of the mass step requires introducing a mass-dependent $\widehat{R_{\rm B}}$. In fact, only 12 percent relative difference 
between $\widehat{R_{\rm B}}$ in the high- and low-mass supernova hosts ($\widehat{R_{\rm B}}\approx 4.4$ for high-mass hosts 
and $\widehat{R_{\rm B}}=3.9$ for low-mass hosts) allows to eliminate the difference between mean Hubble residuals in the 
two mass bins. Thus, in our modelling based solely on supernova data, the required difference is accounted for by a fraction of 
the measured scatter in $R_{\rm B}$ (30 per cent of $\sigma_{R_{\rm B}}$). This finding is qualitatively consistent with \citet{Brout2021} who showed that the mass step 
is primarily driven by Hubble residuals of red supernovae and thus it can attributed to a difference between extinction coefficients 
in the host stellar mass bins \citep[higher extinction in low-mass hosts; see also][]{Wiseman2022}. However, the effect is much more subtle in our case and $\widehat{R_{\rm B}}$ 
values in both mass bins are compatible with a range of typical extinction coefficients measured in the Milky Way \citep{Fitzpatrick2007,Legnardi2023}.

\section{Summary and conclusions}

We have presented a novel Bayesian hierarchical model for constraining the distributions of intrinsic properties of type Ia supernovae and dust properties in their sight lines from supernova light curve parameters. The model incorporates the observationally motivated assumption that there are two distinct supernova populations and a physically motivated {\em ansatz} for the prior distribution of dust reddening. We have fitted the model to SALT2 light curve parameters of type Ia supernovae in the Hubble flow. The main properties of the best fit model and the resulting supernova populations can be summarised as follows:

(1) The model is strongly favoured over its alternative assuming only one supernova population (with $\Delta BIC=-6.4$).

(2) The model discerns two overlapping supernova populations distinguished primarily by the mean stretch parameter: supernovae with low stretch (fast declining light curves) and high stretch (slowly declining light curves).

(3) Supernovae from the identified population of slow decliners (high stretch) appear to be intrinsically bluer (on average $\Delta c_{\rm int}=-0.077\pm0.029$), with two times stronger dust reddening in their sight lines than supernovae from the opposite population.

(4) Extinction in both supernova populations is described by a broad distribution of $R_{\rm B}$ with maximum ($R_{\rm B}\approx 4.1$) coinciding with the average extinction measured in the Milky Way, and dispersion $\sigma_{R_{\rm B}}\approx 0.9$.

(5) The distribution of dust reddening inferred from the supernova data has a distinct maximum and two tails, in contrast to the exponential model commonly employed in the previous studies.

(6) Intrinsic scatter in the supernova Hubble diagrams vanishes and the model provides a complete explanation of the Hubble residuals arising from the commonly used Tripp calibration \citep{Tripp1998}.

The model presented in our work is entirely driven by supernova data. 
The two populations identified by the model are found to be associated with old and young (star forming) stellar populations and thus they closely coincide with analogous supernova populations identified in previous studies using host galaxy information. The young stellar environment appears to be a natural source of higher column densities of dust (reddening) measured in population 2. The apparent correspondence between the supernova populations and underlying stellar environments provides also a simple framework for understanding the origin of the well known stellar mass step correction in the Tripp calibration. The correction results to some extent from a partial separation of supernova populations driven by an abrupt change of mixing the underlying stellar populations at stellar mass $10^{10}M_{\odot}$.

The proposed model can be effectively used to standardise type Ia supernovae for cosmological analyses. In this approach, one can fit a cosmological model simultaneously with the physical properties of type Ia supernovae and extinction in their sight lines. This is the only complete strategy which allows for modelling the expected redshift evolution of the supernova population fractions (in relation to the star formation history) and perhaps cosmic dust, which otherwise could potentially bias cosmological parameters \citep{Rigault2020}. The model will also be indispensable for providing a quantitative and physical explanation of the recently reported anomalously high slope of the apparent colour correction of type Ia supernovae in the calibration sample from the local measurement of the Hubble constant \citep{Wojtak2022}. The measured colour correction is consistent with being entirely due to extinction with Milky Way-like dust properties. This suggests that the calibration sample contains supernovae originating primarily from only one of the two populations, in contrast to the Hubble flow which visibly mixes the two populations. Quantitative analysis of this problem using the model developed in this study and its possible implications for understanding hidden systematic errors or biases in the local measurements of the Hubble constant will be the subject of our forthcoming paper. 

\section*{Acknowledgments}
This work was supported by a VILLUM FONDEN Investigator grant (project number 16599). RW thanks Luca Izzo, Darach Watson, Nandita Khetan, Christa Gall and Charlotte Angus for inspiring discussions. The authors thank the referee for insightful comments.

\section*{Data availability}
No new data were generated or analysed in support of this research.

\bibliography{master}

\appendix
\onecolumn

\section{Posterior probability}

The probability of observing type Ia supernova with light curve parameters $\pmb{\xi}=\{m_{\rm B},x_{1},c\}$ measured with the covariance matrix $\mathsf{C_{\rm obs}}$, a set of latent variables $\pmb{\phi}$ drawn from two distinct supernova populations with relative fractions described by the weight parameter $0<w<1$ is given by
\begin{equation}
p(\{\pmb{\xi},\pmb{\phi}\})=w\mathcal{G}[\pmb{d};\pmb{\xi}(\pmb{\phi}),\mathsf{C_{\rm obs}}]p_{\rm prior 1}(\pmb{\phi})+(1-w)\mathcal{G}[\pmb{d};\pmb{\xi}(\pmb{\phi}),\mathsf{C_{\rm obs}}]p_{\rm prior 2}(\pmb{\phi}),
\end{equation}
where $\mathcal{G}(\pmb{x};\pmb{a},\mathcal{A})$ is a multivariate normal distribution with mean $\pmb{a}$ and covariance $\mathcal{A}$, $p_{\rm prior\,i}(\pmb{\phi})$ are prior 
distributions of latent variables in the two supernova populations ($i=1,2$). The predicted light parameters are related to the latent variables describing intrinsic 
supernova properties (absolute luminosity $M_{B}$, stretch-related parameter $X_{1}$, intrinsic colour $c_{{\rm int}}$), extrinsic properties of dust in supernova host galaxy 
(dust reddening $E(B-V)_{i}$ and extinction coefficient $R_{B\,i}$), distance modulus $\mu$ in the following way:
\begin{align}
    \pmb{\xi} &= \begin{bmatrix}
           M_{B}+\mu-\alpha X_{1}+\beta c_{{\rm int}} +R_{B}E(B-V)\\
           X_{1} \\
           c_{{\rm int}}+E(B-V)
         \end{bmatrix}.
  \end{align}
Adopting Gaussian priors for supernova intrinsic properties, the extinction coefficient and the distance modulus, i.e.
\begin{align}
p_{\rm prior\,i}(M_{B})  &=  \mathcal{G}(M_{B};\widehat{M_{B\,i}},\sigma_{{\rm int}\,i})  \nonumber \\
p_{\rm prior\,i}(X_{1})   & =  \mathcal{G}(X_{1};\widehat{x_{1\,i}},\sigma_{x_{1\,i}})  \nonumber \\
p_{\rm prior\,i}(c_{{\rm int}})   & =  \mathcal{G}(c_{{\rm int}};\widehat{c_{{\rm int}\,i}},\sigma_{c_{{\rm int}\,i}})  \nonumber \\
p_{\rm prior\,i}(R_{B})   & =  \mathcal{G}(R_{B\,i};\widehat{R_{B\,i}},\sigma_{R_{B\,i}})  \nonumber \\
p_{\rm prior\,i}(\mu)   & =  \mathcal{G}(\mu;\widehat{\mu},\sigma_{\mu}), 
\end{align}
and delta distribution for parameters $\alpha_{i}$ and $\beta_{i}$, i.e. $\delta(\alpha_{i})$ and $\delta(\beta_{i})$, 
the posterior probability distribution for hyperparameters $\pmb{\Theta}$ and parameter $w$ can be simplified to the following form
\begin{align}
p(\{\pmb{\Theta},w\}|\pmb{\xi_{\rm obs}})= &\int_{0}^{\infty}w\mathcal{G}[\pmb{\xi_{1}}(\pmb{\Theta},y_{1});\pmb{\xi_{\rm obs}},\mathsf{C_{1}}(\pmb{\Theta},y_{1})]p_{\rm prior\,1}(y_{1}=E(B-V)_{1}/\tau_{1})\textrm{d}y_{1}+\nonumber \\
 & \int_{0}^{\infty}(1-w)\mathcal{G}[\pmb{\xi_{2}}(\pmb{\Theta},y_{2});\pmb{\xi_{\rm obs}},\mathsf{C_{2}}(\pmb{\Theta},y_{2})]p_{\rm prior\,2}(y_{2}=E(B-V)_{2}/\tau_{2})\textrm{d}y_{2},
 \label{post1SN}
\end{align}
where $\{\tau_{1},\tau_{2}\}$ are the scales of $E(B-V)$ distributions in host galaxies of the two supernova populations,
\begin{align}
    \pmb{\xi_{i}} &= \begin{bmatrix}
           \widehat{M}_{B\,i}+\widehat{\mu}-\alpha_{i}\widehat{x}_{1\,i}+\beta_{i}\widehat{c}_{{\rm int}\,i} +\widehat{R}_{B\,i}\tau_{i}y_{i}\\
           \widehat{x}_{1\,i} \\
           \widehat{c}_{{\rm int}\,i}+\tau_{i}y_{i}
         \end{bmatrix},
  \end{align}
and
\begin{align}
    \mathsf{C}_{i} &= \mathsf{C}_{\rm obs} + \begin{bmatrix}
            (\sigma_{{\rm int}\,i}^{2}+\sigma_{\mu}^{2}+\alpha_{i}^{2}\sigma_{x_{1\,i}}^{2}
            +\beta_{i}^{2}\sigma_{c_{{\rm int}\,i}}^{2}+y_{i}^{2}\tau_{i}^{2}\sigma_{R_{B\,i}}^{2}) 
            & -\alpha_{i}\sigma_{x_{1\,i}}^{2} & \beta_{i}\sigma_{c_{{\rm int}\,i}}^{2} \\
            -\alpha_{i}\sigma_{x_{1\,i}}^{2} & \sigma_{x_{1\,i}}^{2} & 0 \\
            \beta_{i}\sigma_{c_{{\rm int}\,i}}^{2} & 0 & \sigma_{c_{{\rm int}\,i}}^{2}
         \end{bmatrix}.
  \end{align}
For the probability distribution of $E(B-V)$ used in this study, i.e.
\begin{equation}
p_{\rm prior\,i}(y_{i}=E(B-V)/\tau_{i})=\frac{y_{i}^{\gamma-1}\exp(-y_{i})}{\Gamma(\gamma)},
\end{equation}
where $\gamma$ is a shape parameter and $\Gamma(x)$ is the gamma function, the integrals in eq.~(\ref{post1SN}) are calculated numerically. The prior probability distribution is truncated at $y_{\rm max\,i}=E(B-V)/\tau_{i}=10$ 
and renormalized accordingly. The posterior 
probability for the entire sample of supernovae is then given by
\begin{equation}
p(\{\pmb{\Theta},w\} |\{ \pmb{\xi_{1}},...,\pmb{\xi_{N}}\})=\prod_{j=1}^{N}p(\{\pmb{\Theta},w\} |\pmb{\xi_{j}}).
\end{equation}
For the analysis of the cosmological sample we use distance modulus $\widehat{\mu}_{j}$ given by the Planck cosmological model at supernova CMB rest frame redshift $z_{j}$. 
With peculiar velocities as the main source of uncertainties in distances moduli we also use
\begin{equation}
\sigma_{\mu\,j}=\frac{5}{\ln10}\frac{\sigma_{v}}{c}\frac{1}{z_{j}}
\end{equation}
with $\sigma_{v}=250$~km~s$^{-1}$ equal to an upper limit of scatter in peculiar velocities with respect to the linear velocity field \citep{Carrick2015}.

\section{Results for Pantheon+ light curve parameters}
\begin{figure*}
	\centering
	\includegraphics[width=\linewidth]{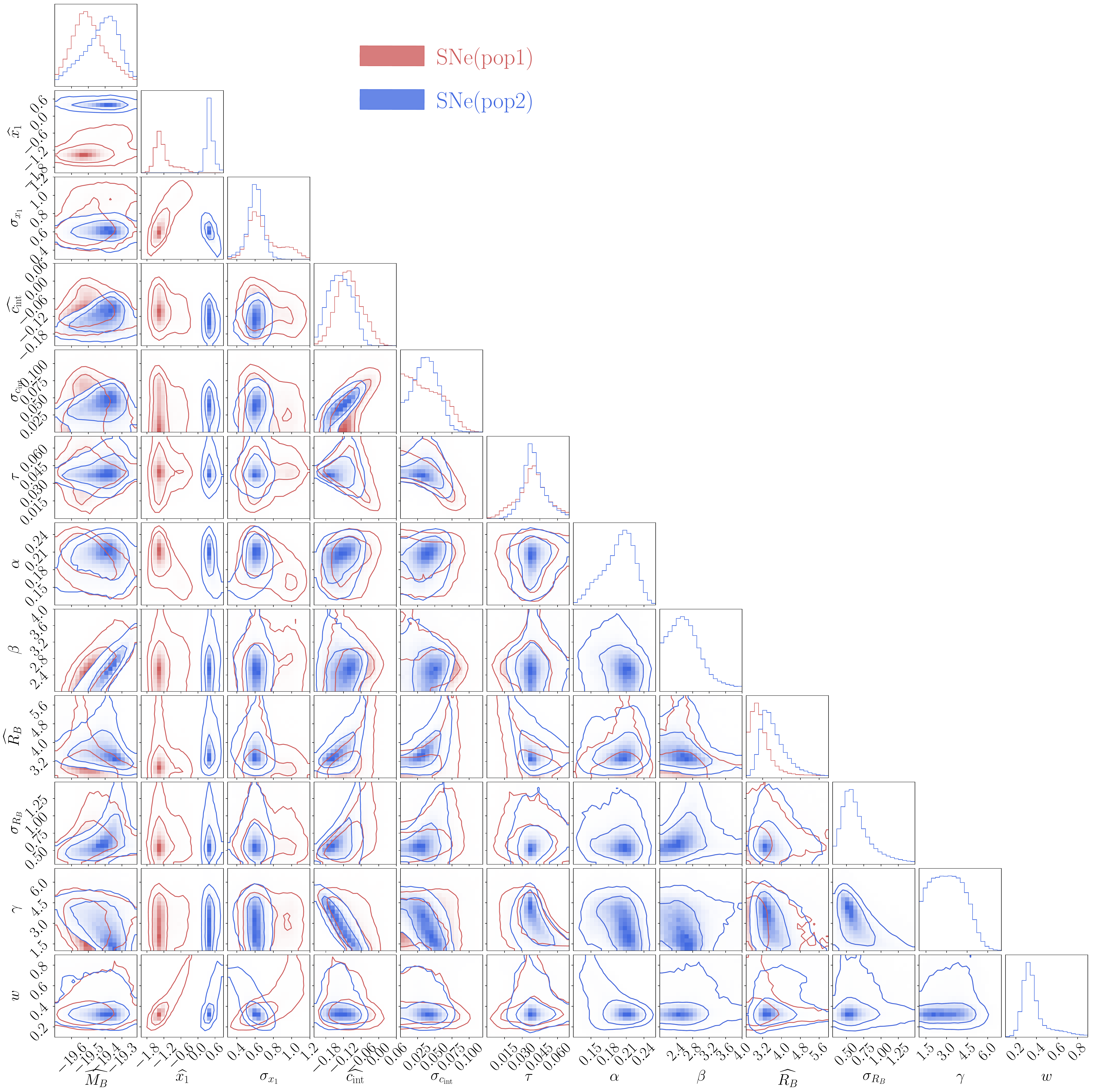}
	\caption{Constraints on hyperparameters of a one-parameter extension of the baseline model (independent mean extinction parameter in both supernova populations) obtained for the 
	main \textit{SuperCal} supernova sample with light curve parameters and redshifts updated from the \textit{Pantheon+} catalogue. The red and blue contours show constraints 
	on parameters in the two supernova populations: population 1 (fast declining supernovae, red) and population 2 (slowly declining supernovae, blue). 
	For the sake of readability, the corner plot is compressed by overlaying the panels with the corresponding sets of hyperparameters. The contours 
	show $1\sigma$ and $2\sigma$ confidence regions containing 68 and 95 per cent of 2-dimensional marginalised probability distributions.
	}
	\label{baseline_pan}
\end{figure*}

Table~\ref{bestmodels_pan} compares best fit parameters of the baseline model with independent $\widehat{R_{\rm B}}$ in both supernova 
populations obtained from the original \textit{SuperCal} supernova sample (left) and its analog with redshifts and light curve parameters updated 
from the \textit{Patheon+} compilation \citep{Brout2022}. Figure~\ref{baseline_pan} shows marginalised posterior distributions for 
the \textit{Patheon+} data.

\begin{table*}
\begin{center}
\begin{tabular}{lcccc}
%\hline
%model & \multicolumn{2}{c}{baseline} & \multicolumn{2}{c}{baseline} \\
\hline
supernova & \multicolumn{2}{c}{SuperCal (156 SNe)} & \multicolumn{2}{c}{Pantheon+ light curve parameters (156 SNe)}  \\
sample & \multicolumn{2}{c}{$0.023<z<0.15$} & \multicolumn{2}{c}{$0.023<z<0.15$}  \\
\hline
 & SNe(pop1) & SNe(pop2) & SNe(pop1) & SNe(pop2) \\
 & fast declining & slowly declining & fast declining & slowly declining \\
 \hline
$\widehat{M_{\rm B}}$ & $ -19.42 ^{+ 0.08 }_{- 0.08 }$ & $ -19.47 ^{+ 0.08 }_{- 0.09 }$ & $-19.48 ^{+ 0.12 }_{- 0.11 }$ & $ -19.41 ^{+ 0.11 }_{- 0.13 }$ \\
& \\
$\widehat{x_{1}}$ & $ -1.08 ^{+ 0.47 }_{- 0.48 }$ & $ 0.50 ^{+ 0.19 }_{- 0.19 }$ & $-1.22 ^{+ 0.31 }_{- 0.26 }$ & $ 0.42 ^{+ 0.11 }_{- 0.11 }$ \\
& \\
$\sigma_{x_{1}}$ & $ 0.80 ^{+ 0.24 }_{- 0.25 }$ & $ 0.55 ^{+ 0.16 }_{- 0.16 }$ & $0.70 ^{+ 0.21 }_{- 0.16 }$ & $ 0.60 ^{+ 0.08 }_{- 0.08 }$ \\
& \\
$\widehat{c_{\rm int}}$ & $ -0.052 ^{+ 0.041 }_{- 0.043 }$ & $ -0.111 ^{+ 0.040 }_{- 0.041 }$ & $-0.101 ^{+ 0.045 }_{- 0.042 }$ & $ -0.132 ^{+ 0.042 }_{- 0.041 }$ \\
& \\
$\sigma_{c_{\rm int}}$ & $ 0.060 ^{+ 0.017 }_{- 0.018 }$ & $ 0.047 ^{+ 0.015 }_{- 0.016 }$ & $0.038 ^{+ 0.028 }_{- 0.028 }$ & $ 0.037 ^{+ 0.018 }_{- 0.019 }$ \\
& \\
$\tau$ & $ 0.022 ^{+ 0.011 }_{- 0.012 }$ & $ 0.033 ^{+ 0.009 }_{- 0.010 }$ & $0.041 ^{+ 0.013 }_{- 0.014 }$ & $ 0.041 ^{+ 0.010 }_{- 0.009 }$ \\
& \\
$\alpha$ & \multicolumn{2}{c}{$ -0.162 ^{+ 0.025 }_{- 0.027 }$} & \multicolumn{2}{c}{$ -0.196 ^{+ 0.029 }_{- 0.026 }$} \\
& \\
$\beta$ & \multicolumn{2}{c}{$ 3.052 ^{+ 0.332 }_{- 0.341 }$} & \multicolumn{2}{c}{$ 2.414 ^{+ 0.559 }_{- 0.687 }$} \\
& \\
$\widehat{R_{\rm B}}$ & $ 3.57 ^{+ 1.19 }_{- 1.01 }$ & $ 4.71 ^{+ 0.83 }_{- 0.74 }$ & $3.09 ^{+ 0.46 }_{- 0.52 }$ & $ 3.69 ^{+ 0.59 }_{- 0.54 }$ \\
& \\
$\sigma_{R_{\rm B}}$ & \multicolumn{2}{c}{$ 0.892 ^{+ 0.362 }_{- 0.328 }$} & \multicolumn{2}{c}{$ 0.636 ^{+ 0.216 }_{- 0.204 }$} \\
& \\
$\gamma$ & \multicolumn{2}{c}{$ 3.429 ^{+ 1.422 }_{- 1.439 }$} & \multicolumn{2}{c}{$ 3.106 ^{+ 1.295 }_{- 1.341 }$} \\
& \\
$w$ & \multicolumn{2}{c}{$ 0.474 ^{+ 0.204 }_{- 0.196 }$} & \multicolumn{2}{c}{$ 0.380 ^{+ 0.102 }_{- 0.102 }$} \\
  \hline
\end{tabular}
\caption{Comparison between best fit model parameters derived from the original \textit{SuperCal} supernova sample 
(left) and its analog with redshifts and light curve parameters updated from the \textit{Patheon+} compilation (right). The adopted 
model allows for independent mean extinction coefficient $\widehat{R_{\rm B}}$ in the two supernova populations. Best fit 
results are provided as the posterior mean values and errors of the credibility range containing 68 per cent of the marginalised 
probabilities (or 68 per cent upper limits).
}
\label{bestmodels_pan}
\end{center}
\end{table*}

\end{document}